\documentclass[11pt,a4paper]{article}   % style for Physical Review B and AJP are similar

\usepackage{amsmath}    % need for subequations
\usepackage{amsfonts}  %note how statements can be commented out
\usepackage{amssymb}
\usepackage{graphicx}   % for figures
\usepackage{subcaption}
\usepackage{mwe}

\usepackage[latin1]{inputenc}     % pour les accents
\usepackage[T1]{fontenc}
\usepackage{authblk} % pour affilation

\usepackage{ulem}    % pour barrer \sout{}
\usepackage{float}   % pour que les images soient inclusent dans le texte
\usepackage{color}   % pour faire des surlignes de couleur
\usepackage{soul}    % pour faire des surlignes de couleur
\graphicspath{{./images/}}
\usepackage[a4paper]{geometry} % pour récupérer les marges
\usepackage{ragged2e}
\justifying

\newtheorem{Theoreme}{THEOREM}
\newtheorem{Definition}{Definition}

\definecolor{lightblue}{rgb}{0.8,0.9,1} % bleu ciel
\begin{document}

\title{The de Broglie-Bohm weak interpretation}
% \affiliation{University Paris Dauphine, Lamsade, 75 016 Paris, France}
\author[1]{Michel Gondran\thanks{michel.gondran@polytechnique.org}}
\author[2]{Alexandre Gondran\thanks{alexandre.gondran@enac.fr}}
\affil[1]{{\small Académie Européenne Interdisciplinaire des Sciences, Paris, France}}
\affil[2]{{\small \'Ecole Nationale de l'Aviation Civile, Toulouse, France}}

\date{}

\maketitle

\begin{abstract}

We define the de Broglie-Bohm (dBB) weak interpretation as the dBB interpretation restricted to particles in unbound states whose wave function is defined in the three-dimensional physical space, and the dBB strong interpretation as the usual dBB interpretation applied to all wave functions, in particular to particles in bound states whose wave function is defined in a $3N$-dimensional configuration space in which $N$ is the number of particules.

We show that the current criticisms of the dBB interpretation do not apply to this weak interpretation and that, furthermore, there are theoritical and experimental reasons to justify the weak dBB interpretation. 

Theoretically, the main reason concern the continuity existing for such particles between quantum mechanics and classical mechanics: we demonstrate in fact that the density and the phase of the wave function of a single-particle (or a set of identical particles without interaction), when the Planck constant tends to 0, converges to the density and the action of a set of unrecognizable prepared classical particles that satisfy the statistical Hamilton-Jacobi equations.  As the Hamilton-Jacobi action pilots the particle in classical mechanics, this continuity naturally concurs with the weak dBB interpretation. 

Experimentally, we show that the measurement results of the main quantum experiments (Young's slits experiment, Stern and Gerlach, EPR-B) are compatible with the de Broglie-Bohm weak interpretation  and everything takes place as if these unbounded particles had trajectories.

In addition, we propose two potential solutions to complete the dBB weak interpretation.

\end{abstract}

\maketitle

%%%%%%%%%%%%%%%%%%%%%%%%%%%%%%%%%%%%%%%%%%%%%%%%%%%%%%%%%%%%%%%%%%%%%%%%%%%%%%
%%%%%%%%%%%%%%%%%%%%%%%%%%%%%%%%%%%%%%%%%%%%%%%%%%%%%%%%%%%%%%%%%%%%%%%%%%%%%%
%%%%%%%%%%%%                    INTRODUCTION                     %%%%%%%%%%%%%
%%%%%%%%%%%%%%%%%%%%%%%%%%%%%%%%%%%%%%%%%%%%%%%%%%%%%%%%%%%%%%%%%%%%%%%%%%%%%%
%%%%%%%%%%%%%%%%%%%%%%%%%%%%%%%%%%%%%%%%%%%%%%%%%%%%%%%%%%%%%%%%%%%%%%%%%%%%%%
\section{Introduction}
\label{sect:intro}

The interpretation of the wave function is always the main problem when it comes to understanding quantum mechanics. Since the 1927 Solvay conference, the de Broglie-Bohm (dBB) pilot-wave, has been 
one of the primary competitor of the Copenhagen interpretation. 
As the recent books by Lalo\"e~\cite{Laloe2012} and Bricmont~\cite{Bricmont2016} recalled, the Copenhagen interpretaion 
presents an inconsistency with the postulate of reduction of the wave packet; moreover, the problem of measurement is not solved even with the theory of decoherence. The main three alternatives are: 
1) to consider the superposition as a physical reality; the option taken by Everett in his multiple world's interpretation; 
2) to modify the Sch\"odinger equation to make it nonlinear and/or non-deterministic; this approach is taken for example by Ghirardi, Rimini and Weber (GRW);
3) to complete quantum mechanics with supplementary variables; this is the approach proposed by the dBB theory that adds positions to the standard formalism of quantum mechanics.
Louis de Broglie first introduced the pilot-wave for single-particle waves (one body systems) in early 1927 \cite{deBroglie1927}, then generalized the pilot-wave for many-particles waves (many-body systems) in 1927 at the Solvay conference \cite{deBroglie1928}.

However in 1928, Louis de Broglie abandoned the pilot wave. David Bohm rediscovered it in 1952, and developed it by introducing spin and studying measurement problems \cite{Bohm1952}. In his 1951 book \cite{Bohm1951}, he defines the EPR-B experiment starting from Einstein, Podolski and Rosen's experiment to replace the measurement of the positions and velocities by the measurement of the spins.

It is this EPR-B experiment related to the Broglie-Bohm interpretation that John Bell used  
to revive the debate on the completeness of standard quantum mechanics with hidden variables. This interpretation has been developed in recent years under the name of Bohmian Mechanics by many authors \cite{Holland1993, Durr1992, Durr2004, Bacciagaluppi2009, Norsen2010}.
Bernard d'Espagnat considers the dBB theory as "an useful theoretical laboratory" as it produces images that are of great help for the imagination~\cite{dEspagnat2003b}.

The purpose of this article is to contribute to the interpretation of quantum mechanics using a rigorous methodology based on the four precepts of the Descates' \textit{Discourse on the Method of Rightly Conducting One's Reason and of Seeking Truth in the Sciences}~\cite{Descartes1637} (1637). In this work, Descartes advises us to apply the following four precepts "\textit{[taking] the firm and unwavering resolution never in a single instance to fail in observing them}": 

"\textit{The first precept was never to accept anything for true which I did not clearly know to be such; that is to say, carefully to avoid precipitancy and prejudice, and to comprise nothing more in my judgment than what was presented to my mind so clearly and distinctly as to exclude all ground of doubt.}

\textit{The second, to divide each of the difficulties under examination into as many parts as possible, and as might be necessary for its adequate solution.}

\textit{The third, to conduct my thoughts in such order that, by commencing with objects the simplest and easiest to know, I might ascend by little and little, and, as it were, step by step, to the knowledge of the more complex; assigning in thought a certain order even to those objects which in their own nature do not stand in a relation of antecedence and sequence.}
        
\textit{And the last, in every case to make enumerations so complete, and reviews so general, that I might be assured that nothing was omitted.}" (Ren\'e Descartes, 1637 \cite{Descartes1637})

 Descartes' first precept encourages us to solve the problem of the interpretation of quantum mechanics by focusing on the dBB theory, but with great caution, avoiding any precipitation in our judgments as long as there is the slightest doubt.
This questioning of the dBB theory will be done by considering the main criticisms that have been levelled at it.

Descartes' second precept cautions us not to take the problem as a whole from the outset, but to divide it into "as many parts as possible, and as might be necessary for its adequate solution." This is the basis of the Cartesian method. However the question of how to divide the problem into subproblems is not a simple one. We follow d'Espagnat's remark regarding "the preparation of systems" in carrying out this decomposition~: 
"\textit{I look at the fundamental physics as it exists today, that of atoms and particles. Going into the details of the mathematical formalism which underlies it, I see it as being entirely based on the notions of "preparation of systems" and "measurement of observables." I occurs to me that these precepts are anthropocentric. I have sought in vain for an approach that replaces them with non-anthrpocentric precepts but must conclude that no convineing attemps has made.}" (Bernard d'Espagnat, 1982 \cite{dEspagnat1982}).
Indeed, the way to prepare a quantum system is the division chosen in the article.

Descartes' thid precept incites us to begin with the simplest systems, and easiest to know, to ascend gradually, as by degrees, to  knowledge of the most complex. And the fourth precept is to be exhaustive in this analysis so as to be "assured of omitting nothing".

In this paper, we will restrict our research on the dBB theory to the simplest objects, i.e. quantum systems prepared in such a way that the main criticisms of the dBB theory do not apply. The other cases corresponding to more complex preparations of quantum systems, linked to Descartes'third and fourth precepts, will be studied in a later paper.

To our knowledge, this approach by decomposition of the quantum systems from their preparation has not been envisaged.

%\bigskip
Let us begin by recalling the main objective criticisms of the dBB interpretation.

\subsection*{Criticisms of the dBB interpretation}
\subsubsection*{$3N$-dimensional configuration space}
From as early as 1927, the first criticism of the de Broglie-Bohm pilot-wave theory concerned the many-particles  wave function in a $3N$-dimensional configuration space (with $N$, the number of particules) which is considered as a fictitious wave in an abstract space, not as a physical wave in a physical space. Even de Broglie, Bohm and Bell have never been completely convinced of the interest of the pilot wave in the configuration space as shown by the following quotations of de Broglie, Bohm, Heisenberg and Bell:

"\textit{It appears to us very probable that the wave}
\begin{equation*}
\Psi=a(q_1, q_2,...,q_n) cos\frac{2\pi}{h}\varphi(t, q_1,...q_n),
\end{equation*}	
\textit{a solution of the Schr\"odinger equation, is only a fictitious wave, which in the \textit{Newtonian approximation}, plays for the representative point of the system in configuration space the same role of pilot wave and of probability wave that the wave $ \Psi$ plays in ordinary space in the case of a single material point}." (Louis de Broglie 1927~\cite{deBroglie1927} cited by Norsen~\cite{Norsen2010})

\textit{"...a serious problem confronts us when we extend the theory ... to the
treatment of more than one electron. This difficulty arises in the circumstance
that, for this case, Schr\"odinger's equation (and also Dirac's
equation) do not describe a wave in ordinary 3-dimensional space, but
instead they describe a wave in an abstract 3N-dimensional space, where
N is the number of particles. While our theory can be extended formally
in a logically consistent way by introducing the concept of a wave in
a 3N-dimensional space, it is evident that this procedure is not really
acceptable in a physical theory, and should at least be regarded as an
artifice that one uses provisionally until one obtains a better theory
in which everything is expressed once more in ordinary 3-dimensional
space."} (Bohm 1987~\cite{Bohm1987}, p. 117  cited by Norsen et al.~\cite{Norsen2014}).

\textit{"For [de Broglie and] Bohm, the particles are "objectively real" structures,
like the point masses of classical mechanics. The waves in configuration space also are objective real fields, like electric fields.... [But]
what does it mean to call waves in configuration space "real"? This space
is a very abstract space. The word "real" goes back to the Latin word
"res", which means "thing"; but things are in the ordinary 3-dimensional
space, not in an abstract configuration space."} (Heisenberg 1955~\cite{Heisenberg1955} cited by Norsen et al.~\cite{Norsen2014}).

\textit{
"Note that in this [theory] the wave is supposed to be just as "real" and
"objective" as say the fields of classical Maxwell theory - although its
action on the particles ... is rather original. No one can understand this
theory until he is willing to think of $\Psi $ as a real objective field rather
than just a "probability amplitude". Even though it propagates not in
3-space but in 3N-space."} (Bell 1994~\cite{Bell1994}, p. 128  cited by Norsen et al.~\cite{Norsen2014}).

\subsubsection*{Zero speed of the electron in the ground state}
The second criticism concerns the zero speed of the electron in the ground state. It is Heisenberg and Pauli's main objective criticism. As Heisenberg wrote in 1958:

\textit{"One consequence of this interpretation, as pointed out by Pauli, is that the electrons in the ground state of many atoms should be at rest, that is to say they have no orbital motion around the nucleus. This seems in contradiction with experiments, because the measurements of the velocities of electrons in the ground state (for example, through the Compton effect) always reveal a distribution of velocities in the ground state, a distribution - in accordance with the rules of quantum mechanics - given by the square of the wave function in momentum or space velocity "}(Heisenberg, 1958, p. 167-168).

\subsubsection*{Indiscernibility and quantization}
The third criticism concern the possibility of reconciliation of behavior between the quantum and classical systems that allows the pilot wave. It is based on differences that seem fundamental: indiscernibility and quantization of energy in quantum mechanics, discernability and continuity of energy in classical mechanics.

\subsubsection*{EPR-B experiment}

Today, the main criticism concerns the interpretation of the EPR-B experiment with the discussion on the non-local hidden variables and on the Bell's inequalities.

We define as \textbf{weak} the de Broglie-Bohm theory restricted to unbounded particles whose wave function is defined in a 3D physical space, and \textbf{strong} the usual de Broglie-Bohm interpretation applied to all wave functions.

In this paper we introduce this concept of de Broglie-Bohm weak theory and we show that it is a scientifically convincing theory that is not contradicted by the four previous criticisms.

The restriction to a wave function in the 3D physical space eliminates the first criticism regarding 3N-dimensional configuration space.

The restriction to a particle in an unbound state nullifies the second criticism on the speed of electrons in ground states. 

We will see that the restriction to a particle in an unboud state also nullifies the third criticism.
Indeed, in this case, it is well known that, for particles in unbound states, the energy spectrum is continuous for both classical particles as quantum particles. We then propose a convergence theorem from quantum mechanics to classical mechanics, when Planck's constant tends to zero, where the phase of the wave function pilots quantum particles in the same vay the Hamilton-Jacobi action pilots classical particles.

We will also see in section V how the fourth criticism about EPR-B experiment can be elimined.

Our article is organised as follows. In section II, we recall how the Hamilton-Jacobi action pilots the particle in classical mechanics. We also introduce the Minplus
path integral, which links the Hamilton-Jacobi and the Euler-Lagrange actions and which provides a clear interpretation of the least action principle. Introducing an initial density to classical particles having an initial Hamilton-Jacobi action, we define the new concept of unrecognizable classical particles.  These classical particles satisfy the statistical Hamilton-Jacobi equations and provide a solution, in classical mechanics, to the Gibbs paradox.

In section III, we show that the density and the phase of the wave function in the 3-dimensional physical space of a set of identical particles without interaction, converge, when the Planck constant $\hbar$  mathematically tends  to 0, to the density and the action of a set of unrecognizable classical particles that satisfy the statistical Hamilton-Jacobi equations.
This continuity with classical mechanics then naturally leads to the de Broglie-Bohm weak theory.

In section IV, we show that measurement results of the main quantum experiments (Young slits, Stern and Gerlach) are compatible with the de Broglie-Bohm weak interpretation  and everything happens as if these particles in an unbound state had trajectories.

The EPR-B experiment is an experiment with two particles in a six-dimensional configuration space. In section V, we show how to extend the de Broglie-Bohm weak interpretation to the particles entangled by the spin as in the EPR-B experiment.

In section VI, we propose two potential solutions on how to complete the de Broglie-Bohm weak  interpretation for particles in bound states whose wave function is defined in a 3N-dimensional configuration space.

\section{The Hamilton-Jacobi action verifies the Minplus Path Integral and pilots classical particles}
\label{sect:HJ}

The theoretical origin of Louis de Broglie' interpretaion was the connection between the phase of the wave function of quantum mechanics with the  Hamilton-Jacobi action of classical mechanics. 

This Hamilton-Jacobi action is presented in the textbooks as "Hamilton's principal function". However,
its initial action $S_0(\textbf{x})$ is ignored in the physical textbooks such as those of Landau\cite{Landau1976} chap.7 § 47 and Goldstein\cite{Goldstein1966} chap.10; it is only known in some mathematical texbooks\cite{Lions1982, Evans1998} for optimal control problems. We shall see that this initial condition $S_0(\textbf{x})$ is essential for the interpretation of the Hamilton-Jacobi action. 

To adress this oversight, we begin by recalling the connection between the Hamilton-Jacobi action and the principle of least action and the Euler-Lagrange action.

Let us consider a system evolving from the position $\textbf{x}_{0}$ at initial time $ t_0=0$ to the position $\textbf{x}$ at
time $t$ where the variable of control \textbf{u}(s) is the
velocity:
\begin{eqnarray}\label{eq:evolution}
\frac{d \textbf{x}\left( s\right) }{ds}=\mathbf{u}(s),\qquad\forall s\in\left[ 0,t\right]\\
\label{eq:condinitiales}
\textbf{x}(0) =\mathbf{x}_{0},\qquad\textbf{x}(t) =\mathbf{x}.
\end{eqnarray}

If $L(\textbf{x},\dot{\textbf{x}},t)$ is the Lagrangian of the
system, when the two positions $\textbf{x}_0$ and $\textbf{x}$ are
given, \textit{the Euler-Lagrange action} $S_{cl}(\mathbf{x},t;
\textbf{x}_0) $ is the function defined by:
\begin{equation}\label{eq:defactioncondit}
S_{cl}(\textbf{x},t;\mathbf{x}_{0})=\min_{\textbf{u}\left(
s\right),0 \leq s\leq t} \int_{0}^{t}L(\textbf{x}(s),
\textbf{u}(s),s)ds,
\end{equation}
where the minimum (or more generally the minimum or the saddle point) is taken on the
controls $\mathbf{u}(s)$, $s\in$ $\left[ 0,t\right]$, with the
state $\textbf{x}(s)$ given by equations (\ref{eq:evolution}) and
(\ref{eq:condinitiales}). This is the principle of least action
defined by Euler~\cite{Euler1744} in 1744 and Lagrang~\cite{Lagrange1888} in 1755. 

For a non-relativistic particle in a linear potential field with
the Lagrangian $L(\mathbf{x},\mathbf{\dot{x}},t)= \frac{1}{2}m
\mathbf{\dot{x}}^2 + \textbf{K}. \textbf{x}$, the Euler-Lagrange action is
equal to 
$
S_{cl}( \mathbf{x},t; \textbf{x}_0)= m
\frac{(\textbf{x}-\textbf{x}_0)^2}{2 t}+ \frac{K .(\textbf{x} +
\textbf{x}_0)}{2}t - \frac{K^2}{24 m}t^3$
and the initial velocity of the trajectory minimizing the action is $
\widetilde{\textbf{v}}_0= \frac{\textbf{x}-\textbf{x}_0}{ t}- \frac{K t}{2 m}$.
Then, $\widetilde{\textbf{v}}_0$ depends on the position $\textbf{x}$ of the particle at the final
time $t$. This dependence on the "final causes" is
general.

One must conclude that, without knowing the initial velocity, the
Euler-Lagrange action answers a problem posed by an observer: "What would the velocity of the particle be at the
initial time to attain $\textbf{x}$ at time $t$?" The resolution
of this problem implies that the observer solves the
Euler-Lagrange equations after the
observation of $\textbf{x}$ at time $t$. This is an \textit{a
posteriori} approach.

The Hamiton-Jacobi action will overcome this \textit{a priori} lack of  knowledge of the initial velocity in the Euler-Lagrange action. Indeed, at the initial time, the Hamilton-Jacobi action $S_0(\textbf{x})$ is known. The knowledge of this initial action $S_0(\textbf{x})$ involves the knowledge of the velocity field at the initial time that satisfies $\textbf{v}_0(\textbf{\textbf{x}})=\dfrac{\nabla S_0(\textbf{x})}{m}$, see equation (\ref{eq:eqvitesse}). \textit{The Hamilton-Jacobi action } $S(\mathbf{x},t) $ at $\textbf{x} $ and time t is then the function defined by:
\begin{equation}\label{eq:defactionHJ}
S(\mathbf{x},t)=\min_{\textbf{x}_0;\mathbf{u}\left( s\right),0
\leq s\leq t }\left\{ S_{0}\left( \mathbf{x}_{0}\right)
+\int_{0}^{t}L(\textbf{x}(s), \mathbf{u}(s),s)ds\right\}
\end{equation}
where the minimum is taken on all initial positions $\textbf{x}_0$
and on the controls $\mathbf{u}(s)$, $s\in$ $\left[ 0,t\right]$,
with the state $\textbf{x}(s)$ given by the equations
(\ref{eq:evolution})(\ref{eq:condinitiales}).

For the non-relativistic Lagrangian $L(\mathbf{x},\mathbf{\dot{x}},t)= \frac{1}{2}m
\mathbf{\dot{x}}^2 - V(\textbf{x})$, we deduce the well-known
result\cite{Evans1998}:

\begin{Theoreme}\label{th:eqactionHJ}- 
The Hamilton-Jacobi action $S\left( \mathbf{x,}t\right) $ is a solution to the Hamilton-Jacobi
equations:
\begin{equation}\label{eq:HJ}
\frac{\partial S}{\partial t}+\frac{1}{2m}(\nabla
S)^{2}+V(\textbf{x},t)=0
\end{equation}
\begin{equation}\label{eq:condinitialHJ}
S(\textbf{x},0)=S_{0}(\textbf{x}).
\end{equation}
and the velocity of a non-relativistic classical particle is given for each point $ \left(
\mathbf{x,}t\right)$ \textit{by}:
\begin{equation}\label{eq:eqvitesse}
\mathbf{v}\left( \mathbf{x,}t\right) =\frac{\mathbf{\nabla }S\left( \mathbf{%
x,}t\right) }{m}
\end{equation}
\end{Theoreme}

The initial condition $S_{0}(\textbf{x}) $ is 
mathematically necessary to obtain the general solution to the
Hamilton-Jacobi equations (\ref{eq:HJ})(\ref{eq:condinitialHJ}). Physically, it is a condition that describes the preparation of the particles. We will see that this
initial condition is the key to understanding the least action principle.

Noting that because $S_0(\textbf{x}_0)$ does not play a role in the minimization on $\textbf{u}(s)$ in
(\ref{eq:defactionHJ}), we
obtain a relation between the Hamilton-Jacobi action and
the Euler-Lagrange action:
\begin{equation}\label{eq:ELHJ}
S(\mathbf{x},t)=\min_{\textbf{x}_0} ( S_{0}\left(
\mathbf{x}_{0}\right) + S_{cl}(\textbf{x},t;\textbf{x}_0) ).
\end{equation}
It is an equation that generalizes the Hopf-Lax and Lax-Oleinik formula,
\cite{Lions1982, Evans1998} $S(\mathbf{x},t)=\min_{\textbf{x}_0} ( S_{0}\left(
\mathbf{x}_{0}\right) + m \dfrac{(x-x_0)^2}{2t})$ that corresponds to the particular case of the free particle where the Euler-Lagrange action is equal to $ m \dfrac{(x-x_0)^2}{2t}$.

Let us note that there exists a new branch of mathematics, \textit{Minplus analysis},
which studies nonlinear problems through a linear approach, cf.
Maslov~\cite{Maslov1992} and
Gondran.~\cite{Gondran1996, Gondran2008} The idea is to
replace the usual scalar product $\int_{X} f(x) g(x) dx$ by the
Minplus scalar product:
\begin{equation}
    (f,g) =\,\underset{x\in X}{\inf }\left\{ f(x)+g(x) \right\}
\end{equation}

In Minplus analysis,
the Hamilton-Jacobi equation is linear, because if
$S_1(\textbf{x},t)$ and $S_2(\textbf{x},t)$ are solutions to
(\ref{eq:HJ}), then $\min\{\lambda + S_1(\textbf{x},t), \mu +
S_2(\textbf{x},t)\}$ is also a solution to the Hamilton-Jacobi
equation (\ref{eq:HJ}).

The Hamilton-Jacobi action $S(\textbf{x},t)$ is then given by the
Minplus integral (\ref{eq:ELHJ}) that we call the \textit{Minplus Path Integral}. 
Indeed, this Minplus integral
(\ref{eq:ELHJ}) for the action in classical mechanics is analogous to the Feynmann path integral for the wave function in quantum mechanics. In the Feynman path integral \cite{Feynman1965} (p. 58),
the wave function $\Psi(\textbf{x},t)$ at time $t$ is written as a
function of the initial wave function $\Psi_{0}(\textbf{x})$:
\begin{equation}\label{eq:interFeynman}
\Psi(\textbf{x},t)= F(t,\hbar) \int \Psi_{0}(\textbf{x}_{0})
\exp\left(\frac{i}{\hbar}S_{cl}(\textbf{x},t;\textbf{x}_{0}\right)
d\textbf{x}_0
\end{equation}
where $F(t,\hbar)$ is an independent function of $\textbf{x}$ and
of $\textbf{x}_{0}$.

Minplus analysis has
many applications in physics: it establishes the correspondence between microscopic
and macroscopic models \cite{Gondran2008}; it is also at the basis of Minplus-wavelets to compute H\"older exponents for fractal\cite{Gondran2014} and multifractal functions.\cite{Gondran2016}

For a particle in a linear potential $V(\textbf{x})= - \textbf{K}
.\textbf{x}$ with the initial action $S_0(\textbf{x})= m
\textbf{v}_0 \cdot \textbf{x}$, we deduce from equation
(\ref{eq:ELHJ}) that the Hamilton-Jacobi action is equal to 
\begin{equation}\label{eq:actlin}
S\left( \textbf{x},t\right)=m \textbf{v}_0 \cdot \textbf{x} -
\frac{1}{2} m \textbf{v}_0^2 t +\textbf{K}.\textbf{x} t -
\frac{1}{2} \textbf{K}.\textbf{v}_{0} t^{2} - \frac{\textbf{K}^2
t^3}{6 m}.
\end{equation}
Equation (\ref{eq:eqvitesse}) shows that the solution $S\left(
\mathbf{x,}t\right) $ to the Hamilton-Jacobi equations yields the
velocity field  for each point ($\textbf{x},t$) from the velocity
field $\frac{\nabla S_0(\textbf{x})}{m} $ at the initial time. In
particular, if at the initial time, we know the initial position
$\textbf{x}_{init}$ of a particle, its velocity at this time is
equal to $\frac{\nabla S_0(\textbf{x}_{init})}{m}$. From the
solution $ S\left( \mathbf{x,}t\right)$ to the Hamilton-Jacobi
equations, we deduce with (\ref{eq:eqvitesse}) the trajectories of
the particle. The Hamilton-Jacobi action $S\left(
\mathbf{x,}t\right)$ is then a field that "pilots" the particle.

In classical mechanics, we can therefore use both the Hamilton-Jacobi action (5) (6) and the Newton equations, the velocity defined by the equation (7) forming the bond; in classical mechanics there is a duality between the Hamilton-Jacobi action and a classical particle satisfying Newton's equations. We refer to this duality in classical mechanics as the duality action-particle. 

\begin{figure}[h] 
\centering
\includegraphics[width=0.7\linewidth]{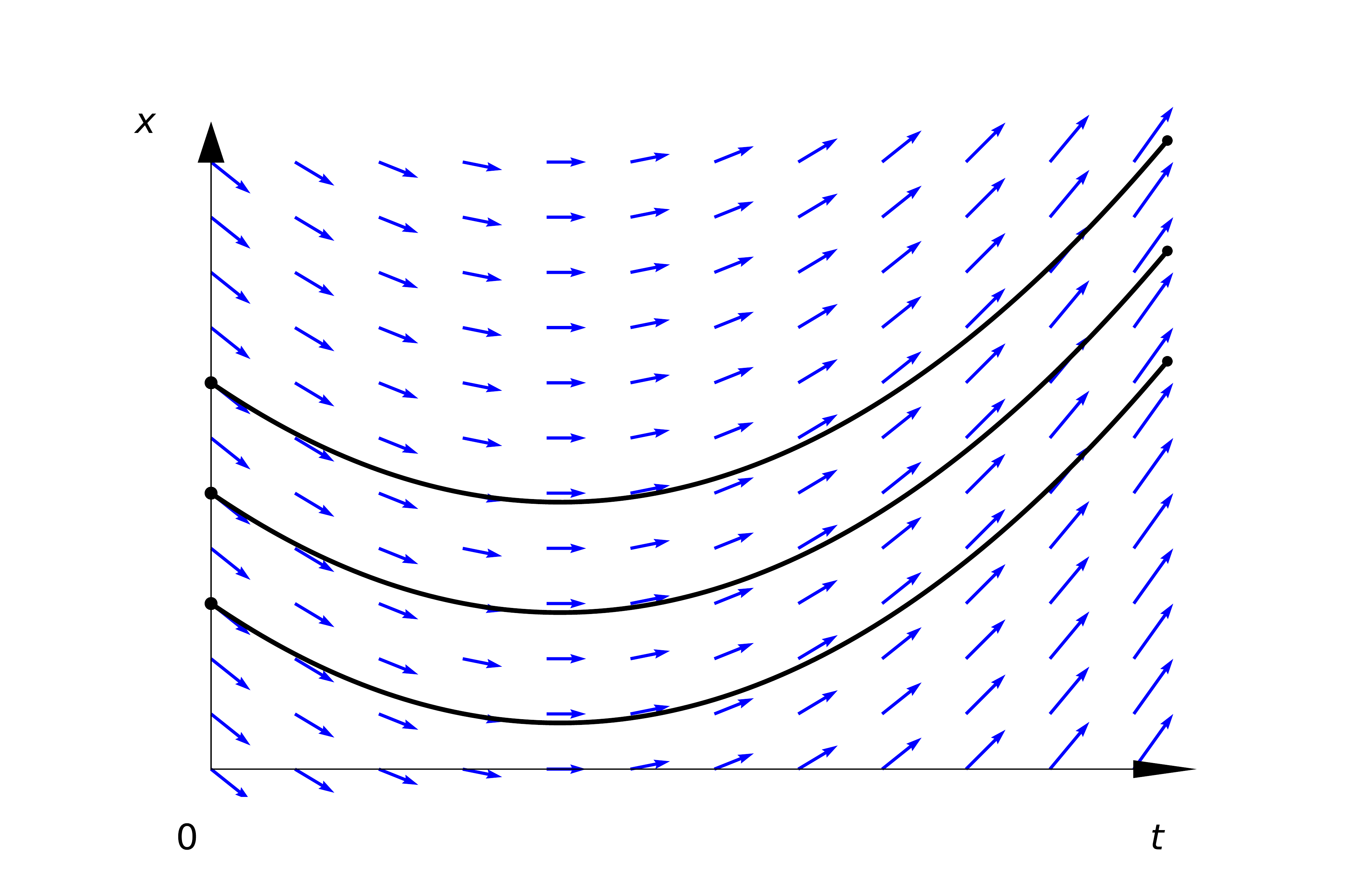}
\caption{\label{fig:fieldHJ} Velocity field (blue arrows) that corresponds to the Hamilton-Jacobi action $ S\left( \mathbf{x},t\right)=m \textbf{v}_0 \cdot
\textbf{x} - \frac{1}{2} m \textbf{v}_0^2 t +\textbf{K}.\textbf{x}
t - \frac{1}{2} \textbf{K}.\textbf{v}_{0} t^{2} -
\frac{\textbf{K}^2 t^3}{6 m}$ ($\textbf{v}(\textbf{x},t)= \frac{\mathbf{\nabla }S\left( \mathbf{x},t\right)}{m}= \textbf{v}_0+ \frac{\textbf{K} t}{m}$) and three trajectories of particles piloted by this field (black lines).}
\end{figure}

For a particle of the previous example, the initial velocity field is constant, 
$\textbf{v}(\textbf{x},0)= \frac{\mathbf{\nabla }S_{0}\left( \mathbf{x}\right)}{m}= \textbf{v}_0$ 
and the velocity field at time $t$ is also constant, 
$\textbf{v}(\textbf{x},t)= \frac{\mathbf{\nabla }S\left( \mathbf{x},t\right)}{m}= \textbf{v}_0+ \frac{\textbf{K} t}{m}$. 
Figure \ref{fig:fieldHJ} shows these velocity fields.

Equations (\ref{eq:HJ}), 
(\ref{eq:condinitialHJ}) and (\ref{eq:eqvitesse}) show
that the Hamilton-Jacobi action $S(\mathbf{x},t)$ not only solves a given problem with a single initial condition $\left(
\mathbf{x}_{0}, \frac{\mathbf{\nabla }S_{0}\left(
\mathbf{x}_{0}\right) }{m}\right) $, but also a set of problems with an
infinity of initial
conditions, all the pairs $\left( \mathbf{y},%
\frac{\mathbf{\nabla }S_{0}\left( \mathbf{y}\right) }{m}\right)$.
It answers the following question: "If we know the action (or the
velocity field) at the initial time, can we determine the action
(or the velocity field) at each later time?" This problem is
solved sequentially by the (local) evolution equation
(\ref{eq:HJ}). This is an \textit{a priori} point of view. It is the
problem solved by Nature with the principle of least action.

Minplus analysis can help to explain the mathematical differences between Hamilton-Jacobi and Euler-Lagrange actions: \textit{The  Euler-Lagrange action
$S_{cl}(\textbf{x},t;\textbf{x}_0)$ is the elementary solution in the Minplus analysis to
the Hamilton-Jacobi equations
(\ref{eq:HJ})(\ref{eq:condinitialHJ}) with
the initial condition}
\begin{equation}\label{eq:conditinitEL}
S_0(\mathbf{x})= \delta_{\min}(\textbf{x}- \textbf{x}_0)= \left\{\begin{array}{ll}
                  0&\textmd{if}\quad \mathbf{x}=\mathbf{x}_0,\\
		  +\infty&\textmd{otherwise}
                 \end{array}
\right.
\end{equation}
where $\delta_{\min}(\textbf{x})$ is the analog to the Dirac distribution $\delta(\textbf{x})$ in Minplus analysis.

In classical mechanics, a particle is usually considered as a point and is described by its mass $m$, 
its charge if it has one, as well as its position $\textbf{x}_0$ and velocity $\textbf{v}_0$ at the initial instant. 
If the particle is subject to a potential field $V (\textbf{x})$, we can deduce its path because its future 
evolution is given by Newton's or Lorentz's equations. This is why classical particles  are considered 
distinguishable. We will show, however, that a classical particle can be considered unrecognizable depending on how it is prepared.

We now consider a particle within a stationary beam of classical
identical particles. For a particle of this beam, we do not know at the initial instant the exact position 
or the exact velocity, only the characteristics describing the beam, that is to say, 
an initial probability density  $\rho _{0}\left(\mathbf{x}\right)$ and an initial velocity 
field $\mathbf{v}_0(\textbf{x})$ known through the initial action $S_0(\textbf{x})$ by the equation
$\mathbf{v}_0(\textbf{x})= \frac{\nabla S_0(\textbf{x})}{m}$ where
$m$ is the particle mass. This yields the following definition:

\begin{Definition}[Unrecognizable classical particle]\label{defparticulesindiscernedenmc}- 
A classical particle is said to be unrecognizable prepared 
when only the characteristics of the beam from which it comes (initial probability density 
$\rho _{0}\left( \textbf{x}\right)$ and initial action $S_0(\textbf{x}) $) are defined at the initial time.
\end{Definition}

In contrast, we have:

\begin{Definition}[Recognizable classical particle]\label{defparticulesdiscernedenmc}- 
A classical particle is said to be recognizable prepared, if one knows, 
at the initial time, its position $\textbf{x}_0 $ and velocity $\textbf{v}_0 $.
\end{Definition}

The notion of unrecognizability we have introduced does not depend on the observer's knowledge, 
but is related to the mode of preparation of the particle. 
 
Let us consider $N$ unrecognizable particles, that is to say $N$ identical particles prepared 
in the same way, each with the same initial density $ \rho_{0}\left(\textbf{x}\right)$ and 
the same initial action $S_0(\textbf{x})$, subject to the same potential field $V(\textbf{x})$ 
and which will be caracterized by independent behaviors. This is particularly the case of identical classical 
particles without mutual interaction and prepared the same way. 

We called these particles unrecognizable, and not indistinguishable, because if we knew their initial 
positions, their trajectories would be known. 

This means that the unrecognizable particules can be distinguishable.  
Furthermore, in their enumerations unrecognizable particules have the same 
properties that are usually granted to indistinguishable particles. Thus, if we select $N$ identical 
particles at random from the initial density $\rho_{0}\left( \textbf{x}\right)$, 
the various permutations of the $N$ particles are strictly equivalent 
and correspond, as for indistinguishable particles, to only one configuration. In this framework, 
the Gibbs paradox is no longer paradoxical as it applies to $N$ unrecognizable particles whose different 
permutations correspond to the same configuration as for indistinguishable
particles. This means that if $X$ is the coordinate space of an
unrecognizable particle, the true configuration space of $N$
unrecognizable particles is not $X^N$ but rather $X^N/S_N $
where $S_N$ is the permutation group.

For unrecognizable particles, this yields the following theorem:

\begin{Theoreme}\label{th:eqstatHJ}- The probability density
$\rho \left( \mathbf{x},t\right)$ and the action $S\left(
\mathbf{x,}t\right)$ of classical particles prepared in the same
way, with initial density $\rho_0( \textbf{x})$, with the same
initial action $S_0(\textbf{x})$, and evolving in the same
potential $V(\textbf{x})$, are solutions to the
\textbf{statistical Hamilton-Jacobi equations}:
\begin{eqnarray}\label{eq:statHJ1b}
\frac{\partial S\left(\textbf{x},t\right) }{\partial
t}+\frac{1}{2m}(\nabla S(\textbf{x},t) )^{2}+V(\textbf{x})=0\\
\label{eq:statHJ2b}
S(\textbf{x},0)=S_{0}(\textbf{x})\\
\label{eq:statHJ3b}
\frac{\partial \mathcal{\rho }\left(\textbf{x},t\right) }{\partial
t}+ div \left( \rho \left( \textbf{x},t\right) \frac{\nabla
S\left( \textbf{x},t\right) }{m}\right) =0\\
\label{eq:statHJ4b}
\rho(\mathbf{x},0)=\rho_{0}(\mathbf{x}).
\end{eqnarray}
\end{Theoreme}

Let us apply this theorem to the particular case of a set of classical particles prepared in the same way, with initial density $\rho_0( \textbf{x})= ( 2\pi \sigma_{0} ^{2}) ^{-\frac{3}{2}}e^{-\frac{\left( \textbf{x}-\xi_{0}\right) ^{2}}{2\sigma_{0} ^{2} }}$, with the same
initial action $S_0(\textbf{\textbf{x}})= m \textbf{v}_0 . \textbf{x}$, and evolving in the same linear potential $V(\textbf{x})= - \textbf{K }. \textbf{x}$. Then the Hamilton-Jacobi action $ S(\textbf{x},t)$ is obtained by equation (\ref{eq:actlin}) and the density $\rho(\mathbf{x},t)$ is equal to:
\begin{equation}\label{eq:densit}
\rho(\textbf{x},t)=( 2\pi \sigma_{0} ^{2}) ^{-\frac{3}{2}}e^{-%
\frac{\left( \textbf{x}-\xi_{0}-\textbf{v}_{0}t -\textbf{K}
\frac{t^{2}}{2 m}\right) ^{2}}{2\sigma_{0} ^{2} }}
\end{equation}

\section{Quantum-classical theoretical continuity}
\label{sect:ELHJ}

In this section let us consider a class of wave functions for which we can mathematically show the continuity between classical mechanics and quantum mechanics when the Planck constant tends to 0.
Let us consider the case where the wave function $\Psi(\textbf{x},t)$ is a solution to the Schr\"odinger
equation:
\begin{eqnarray}\label{eq:schrodinger1}
i\hbar \frac{\partial \Psi }{\partial t}=\mathcal{-}\frac{\hbar ^{2}}{2m}%
\triangle \Psi +V(\mathbf{x})\Psi\\
\label{eq:schrodinger2}
\Psi (\mathbf{x},0)=\Psi_{0}(\mathbf{x}).
\end{eqnarray}
With the variable change $\Psi(\mathbf{x},t)=\sqrt{\rho^{\hbar}(\mathbf{x},t)} \exp(i\frac{S^{\hbar}(\textbf{x},t)}{\hbar})$, the Schr\"odinger
equation can be decomposed into Madelung
equations~\cite{Madelung1926} (1926):
\begin{equation}\label{eq:Madelung1}
\frac{\partial S^{\hbar}(\mathbf{x},t)}{\partial t}+\frac{1}{2m}
(\nabla S^{\hbar}(\mathbf{x},t))^2 +
V(\mathbf{x},t)-\frac{\hbar^2}{2m}\frac{\triangle
\sqrt{\rho^{\hbar}(\mathbf{x},t)}}{\sqrt{\rho^{\hbar}(\mathbf{x},t)}}=0
\end{equation}
\begin{equation}\label{eq:Madelung2}
\frac{\partial \rho^{\hbar}(\mathbf{x},t)}{\partial t}+ div
\left(\rho^{\hbar}(\mathbf{x},t) \frac{\nabla
S^{\hbar}(\mathbf{x},t)}{m}\right)=0
\end{equation}
with initial conditions:
\begin{equation}\label{eq:Madelung3}
\rho^{\hbar}(\mathbf{x},0)=\rho^{\hbar}_{0}(\mathbf{x}) \qquad \text{and}
\qquad S^{\hbar}(\mathbf{x},0)=S^{\hbar}_{0}(\mathbf{x}) .
\end{equation}
We consider the following preparation of the
particles~\cite{Gondran2011a,Gondran2012a}.

\begin{Definition}[Unrecognizable quantum particle]\label{defUnrecognizableParticle}
 A quantum particle is said to be unrecognizable
prepared if its initial probability density $\rho^{\hbar}_{0}(\mathbf{x})$
and its initial action $S^{\hbar}_{0}(\mathbf{x})$ are 
functions $\rho_{0}(\mathbf{x})$ and $S_{0}(\mathbf{x})$ not
depending on $\hbar$.
\end{Definition}

It is the case of a set of non-interacting particles all prepared
in the same way: a free particle beam in a linear potential, an
electronic or $C_{60}$ beam in the Young's slits diffraction, or
an atomic beam in the Stern and Gerlach experiment.

\begin{Theoreme}~\cite{Gondran2011a,Gondran2012a} For unrecognizable
prepared quantum particles,
the probability density $\rho^{\hbar}(\textbf{x},t)$ and the
action $S^{\hbar}(\textbf{x},t)$, solutions to the Madelung
equations
(\ref{eq:Madelung1})(\ref{eq:Madelung2})(\ref{eq:Madelung3}),
converge, when $\hbar\to 0$, to the classical density
$\rho(\textbf{x},t)$ and the classical action $S(\textbf{x},t)$,
solutions to the statistical Hamilton-Jacobi equations (\ref{eq:statHJ1b})(\ref{eq:statHJ2b})(\ref{eq:statHJ3b})(\ref{eq:statHJ4b}). 
\end{Theoreme}

We give some indications on the demonstration of this theorem and then
we propose an interpretation. Let us consider the case where the
wave function $\Psi(\textbf{x},t)$ at time $t$ is written as a
function of the initial wave function $\Psi_{0}(\textbf{x})$ by
the Feynman paths integral \cite{Feynman1965} (\ref{eq:interFeynman}). For an  unrecognizable
prepared quantum particle, the wave function is written
$ \Psi(\textbf{x},t)= F(t,\hbar)\int\sqrt{\rho_0(\mathbf{x}_0)}
\exp(\frac{i}{\hbar}( S_0(\textbf{x}_0)+
S_{cl}(\textbf{x},t;\textbf{x}_{0})) d\textbf{x}_0$. The theorem
of the stationary phase shows that, if $\hbar$ tends towards 0, we
obtain $ \Psi(\textbf{x},t)\sim
\exp(\frac{i}{\hbar}min_{\textbf{x}_0}( S_0(\textbf{x}_0)+
S_{cl}(\textbf{x},t;\textbf{x}_{0}))$, that is to say that the
quantum action $S^{h}(\textbf{x},t)$ converges to the function
\begin{equation}\label{eq:solHJminplus}
S(\textbf{x},t)=min_{\textbf{x}_0}( S_0(\textbf{x}_0)+
S_{cl}(\textbf{x},t;\textbf{x}_{0}))
\end{equation}
which is the solution to the Hamilton-Jacobi equation
(\ref{eq:statHJ1b}) with the initial condition (\ref{eq:statHJ2b}).
Moreover, as the quantum density $\rho^{h}(\textbf{x},t)$
satisfies the continuity equation (\ref{eq:Madelung2}), we deduce,
since $S^{h}(\textbf{x},t)$ tends towards $S(\textbf{x},t)$, that
$\rho^{h}(x,t)$ converges to the classical density
$\rho(\textbf{x},t)$, which satisfies the continuity equation
(\ref{eq:statHJ3b}). We obtain both announced convergences.

Before interpreting the result of this theorem, let us consider a particular case: the case of a quantum particle or a set of quantum particles prepared with the initial probability density $\rho_0(\textbf{x})= ( 2\pi \sigma_{0} ^{2}) ^{-\frac{3}{2}}e^{-\frac{\left( \textbf{x}-\xi_{0}\right) ^{2}}{2\sigma_{0} ^{2} }}$ and the initial action $S_0(\textbf{x})= m  \textbf{v}_0 . \textbf{x}$ in a linear potential $V(\textbf{x})= - \textbf{K}. \textbf{x}$.
The resolution \cite{Cohen1977} of the Schr\"odinger equation yields:
\begin{equation}\label{eq:densite}
\rho^{\hbar}(\textbf{x},t)=( 2\pi \sigma_{\hbar} ^{2}( t))
^{-\frac{3}{2}}e^{- \frac{\left(
\textbf{x}-\xi_{0}-\textbf{v}_{0}t-\textbf{K} \frac{t^{2}}{2
m}\right) ^{2}}{2\sigma_{\hbar} ^{2}( t) }}
\end{equation}{\footnotesize
\begin{equation}
S^{\hbar}(\textbf{x},t)=-\frac{3 \hbar }{2}tg^{-1}( \hbar t/2m
\sigma _{0}^{2}) - \frac{1}{2}m\textbf{v}_{0}^{2}t+ m
\textbf{v}_{0} . \textbf{x}+\textbf{K}.\textbf{x} t - \frac{1}{2}
\textbf{K}.\textbf{v}_{0} t^{2} - \frac{\textbf{K}^2 t^3}{6
m}+\frac{\left( \textbf{x}-\xi_{0}-\textbf{v}_{0}t -\textbf{K}
\frac{t^{2}}{2 m}\right) ^{2}\hbar ^{2}t}{8m\sigma
_{0}^{2}\sigma_{\hbar} ^{2}\left( t\right) }
\end{equation}}
with $
\sigma_{\hbar} \left( t\right) =\sigma _{0}\left( 1+\left( \hbar
t/2m\sigma _{0}^{2}\right) ^{2}\right) ^{\frac{1}{2}}$.
When $\hbar \rightarrow 0$, $\sigma_{\hbar} \left( t\right)$ converges to $\sigma_0$, 
$\rho^{\hbar}(\textbf{x},t)$ converges to $\rho(\textbf{x},t) $ defined by (\ref{eq:densit}) and $S^{\hbar}(\textbf{x},t)$ converges to the Hamilton-Jacobi action defined by (\ref{eq:actlin}).

For an unrecognizable quantum particle, the Madelung equations converge to 
the statistical Hamilton-Jacobi equations, which correspond to unrecognizable
prepared
classical particles. We now use the interpretation of the statistical Hamilton-Jacobi 
equations to deduce the interpretation of the Madelung equations. For these unrecognizable
prepared 
classical particles, the density and the action are not sufficient to describe a classical particle. 
It is necessary to know its initial position to deduce its position at time $t$. 
It is logical to do the same in quantum mechanics since there is a convergence of equations. We conclude that an \textit{unrecognizable prepared quantum particle} is not completely described by
its wave function. It is necessary  to add its initial
position and it becomes natural to introduce the de Broglie-Bohm weak interpretation.

In this interpretation, the two first postulates of quantum mechanics, describing the
quantum state and its evolution, must be completed. 
At initial time $t=0$, the state of the particle is
given by the initial wave function $ \Psi_{0}(\textbf{x})$ (a wave
packet) and its initial position $\textbf{X}(0)$; it is the first new postulate of quantum mechanics.
The second new postulate gives the evolution on the wave function and on the position. 
For a single, spin-less particle in a potential
$V(\textbf{x})$, the evolution of the wave function is given by the usual
Schr\"odinger equation (\ref{eq:schrodinger1})(\ref{eq:schrodinger2}) and the evolution of the particle position is given by
\begin{equation}\label{eq:champvitesse}
\frac{d
\textbf{X}(t)}{dt}=\frac{1}{m}\nabla
S^{\hbar}(\textbf{x},t)|_{\textbf{x}=\textbf{X}(t)}.
\end{equation}

In the case of a particle with spin, as in the Stern and Gerlach and EPR-B
experiments, the Schr\"odinger equation must be replaced by the
Pauli or Dirac equations.

The third postulate of quantum mechanics which describes the measurement 
operator (the observable) can be kept. But the three postulates of 
measurement are not necessary: the postulate of quantization, the Born 
postulate of probabilistic interpretation of the wave function and the 
postulate of the reduction of the wave function. As we demonstrate in the following,
these postulates of measurement can be explained~\cite{Gondran2014} on each
example.

We replace these three postulates by a single one, the "quantum equilibrium 
hypothesis", that describes the interaction 
between the initial wave function $\Psi_0(\textbf{x})$ and the initial particle position
$\textbf{X}(0)$: 
For a set of identically prepared particles having $t=0$ 
wave function $\Psi_0(\textbf{x})$, it is assumed that
the initial particle positions $\textbf{X}(0)$ are distributed
according to:
\begin{equation}\label{eq:quantumequi}
P[\textbf{X}(0)=\textbf{x}]\equiv
|\Psi_0(\textbf{x})|^2 =\rho^{h}_0(\textbf{x}).
\end{equation}
This is the Born rule at the initial time.

Then, the probability distribution ($ P[\textbf{X}(t)=\textbf{x}]$) for a set of particles moving
with the velocity field $\textbf{v}^{h}(\textbf{x},t)=\frac{\nabla
S^{h}(\textbf{x},t)}{m}$ satisfies the property of the "equivariance" of the
$|\Psi(\textbf{x},t)|^2$ probability distribution:~\cite{Durr1992}
\begin{equation}\label{eq:quantumequit}
P[\textbf{X}(t)=\textbf{x}] = |\Psi(\textbf{x},t)|^2 =\rho^{h}(\textbf{x},t).
\end{equation}
This is the Born rule at time t.

Then, the de Broglie-Bohm weak interpretation is based on a continuity between classical and quantum mechanics where the quantum particles are statistical prepared with an initial probability densitiy that satisfies the "quantum equilibrium 
hypothesis" (\ref{eq:quantumequi}). It is the case of the three experiments we will study: double-slit and Stern-Gerlach in section IV and EPR-B in section V.

We will revisit these three measurement experiments through
mathematical calculations and numerical simulations.

For each one,
we present the statistical interpretation that is common to the
Copenhagen interpretation and the de Broglie-Bohm pilot wave, then
the trajectories specific to the de Broglie-Bohm weak interpretation.
We show that the precise definition of the initial conditions,
i.e. the preparation of the particles, plays a fundamental
methodological role.

\section{The measurement in the de Broglie-Bohm weak interpretation}
\label{sect:Experiment}

Experimentally, there is a big difference between the unbounded particles in a beam of particles and bounded particles inside atoms or molecules. Indeed, for the unbounded particles in a beam, it is the position of each particle that is measured directly (and not the wave function). The wave function density is obtained only statistically. It is on such particles (scattering states) that Born valided his  wave function statistical interpretation~\cite{Born1926} in 1926.
For the bounded particles inside atoms or molecules, there is not direct measurement of the position inside atoms or molecules, but only a measurement outside when the particles  are extracted from atoms or molecules.

\begin{figure}[h]
\centering
\includegraphics[width=0.9\linewidth]{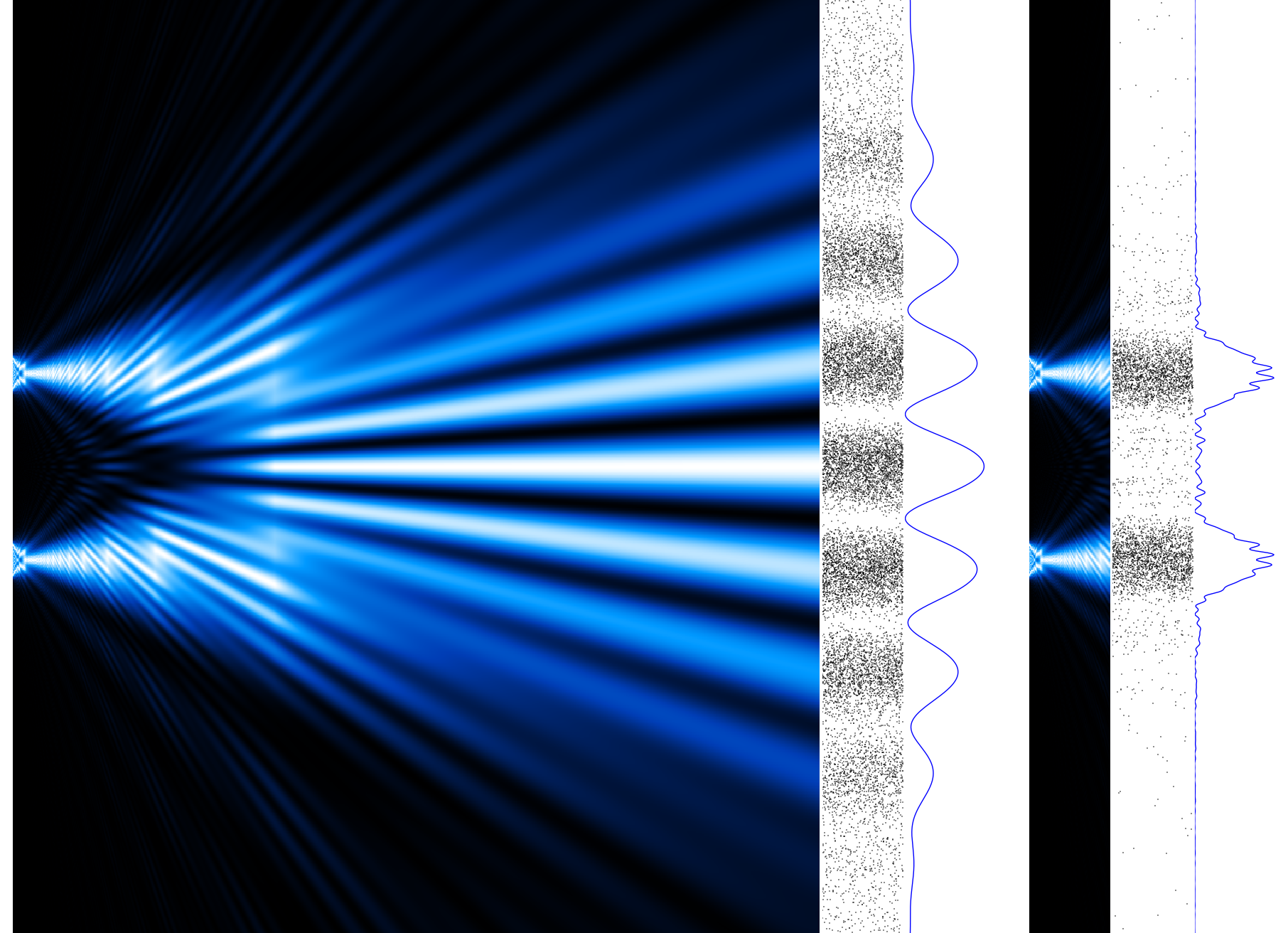}
\caption{\label{fig:traj-Young1}{\small J\"onsson experiment:
evolution of the probability distribution of electrons in the first 10 centimeters after the slits (left figure) and in the first centimeter after the slits (right figure). Impacts of particles and density are also given at 10 cm (left figure) and at 1 cm (right figure).}}
% probability distribution of electrons after the slits and particle impacts at 10cm (left figure) and 1cm (right figure).}
\end{figure}

Because the position is the measured variable for unbounded particles, there is a scientific reason to consider the impacts of these particles on a screen as the positions at the time of impact. For these unbounded particles, it is then natural, experimentally, to introduce the de Broglie-Bohm weak interpretation.

We confirm this remark about the existence of the particle position at moment of measurement on the main quantum experiments (Young slits, Stern and Gerlach).

Figure~\ref{fig:traj-Young1} shows a simulation of the probability distribution in
 J\"onsson's double slit experiment\cite{Jonsson1961} where an electron 
gun emits electrons one by one through a hole with a radius of a few micrometers. 
The electrons, similarly prepared , are represented by the same initial wave function, 
but not by the same initial position. 

\begin{figure}[h]
\begin{center}
\includegraphics[width=0.9\linewidth]{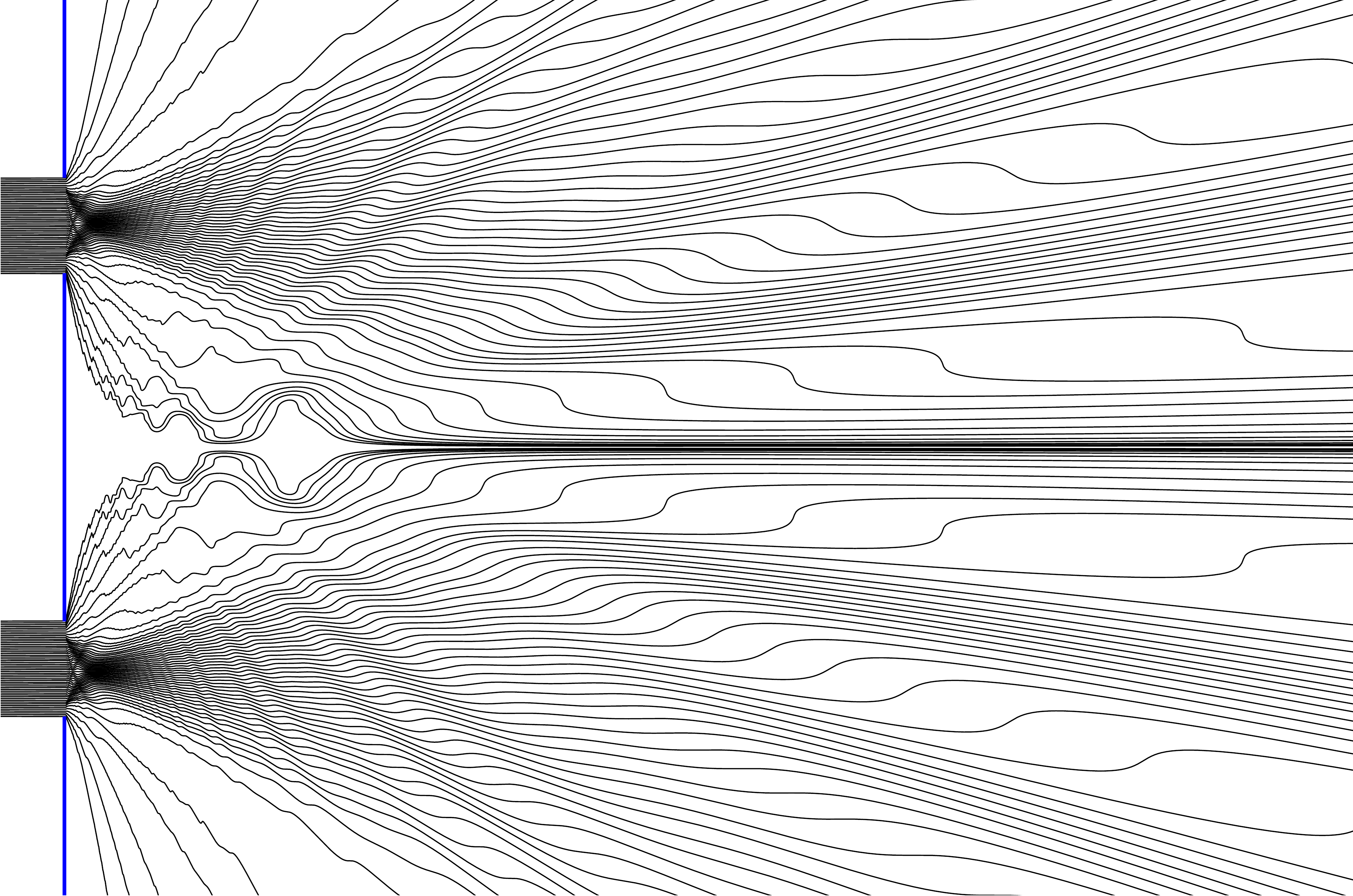}
\caption{\label{fig:traj-Young2} 100 electron trajectories passing through the double-slits for the J\"onsson experiment.} 
\end{center}
\end{figure}

Figure~\ref{fig:traj-Young2} shows a simulation of 100 de Broglie-Bohm trajectories passing through the double-slits.
In the simulation, these initial positions are randomly selected in the
initial wave packet. We have represented only the quantum trajectories through one of two slits.

\begin{figure}[h]
	\centering
	\begin{subfigure}[b]{0.48\textwidth}
		\centering
		\includegraphics[width=\textwidth]{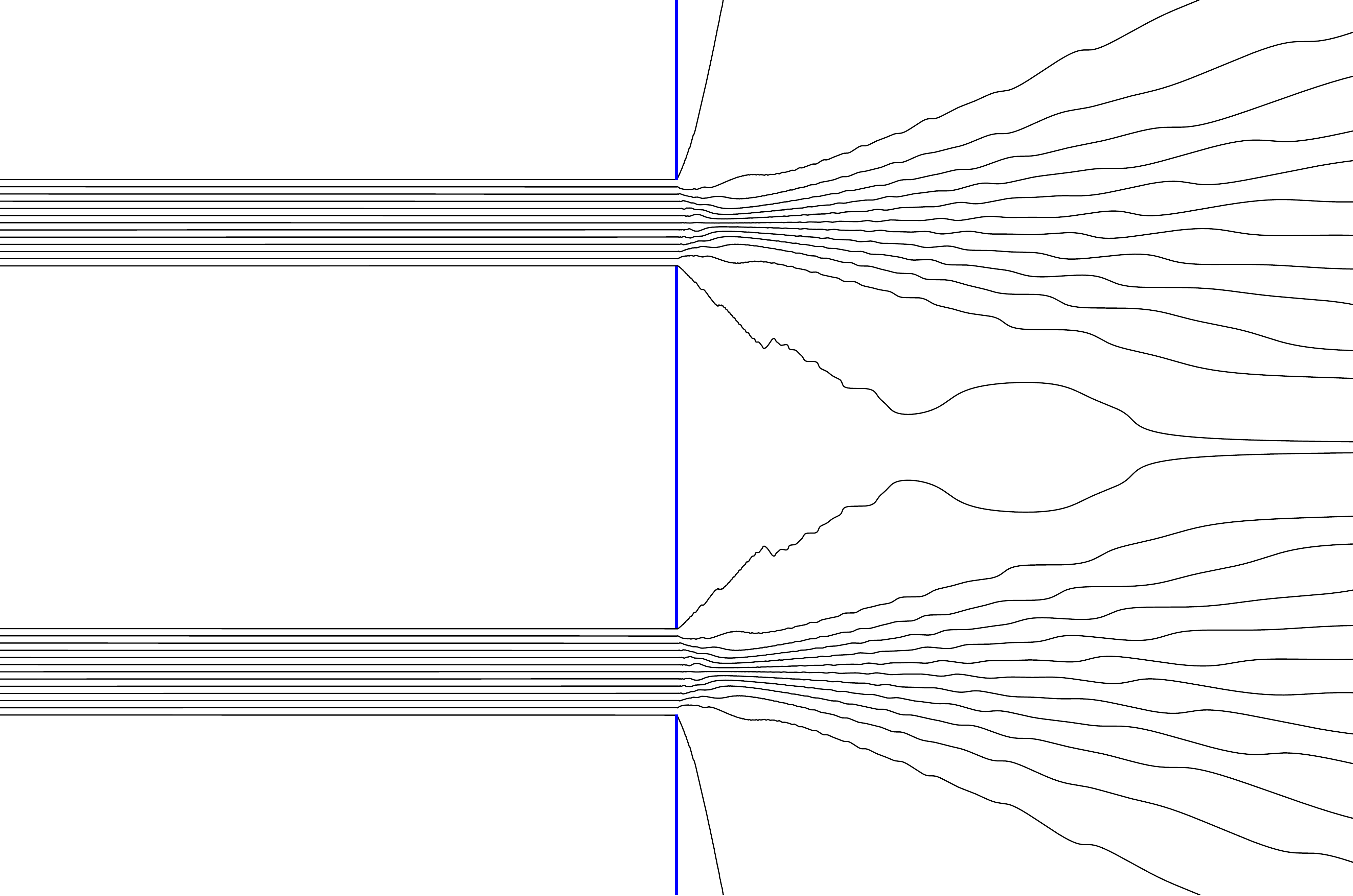}
		\caption[]{{\small $\hbar/10$}}    
		%\label{fig:mean and std of net14}
	\end{subfigure}
	\quad
	%\hfill
	\begin{subfigure}[b]{0.48\textwidth}  
		\centering 
		\includegraphics[width=\textwidth]{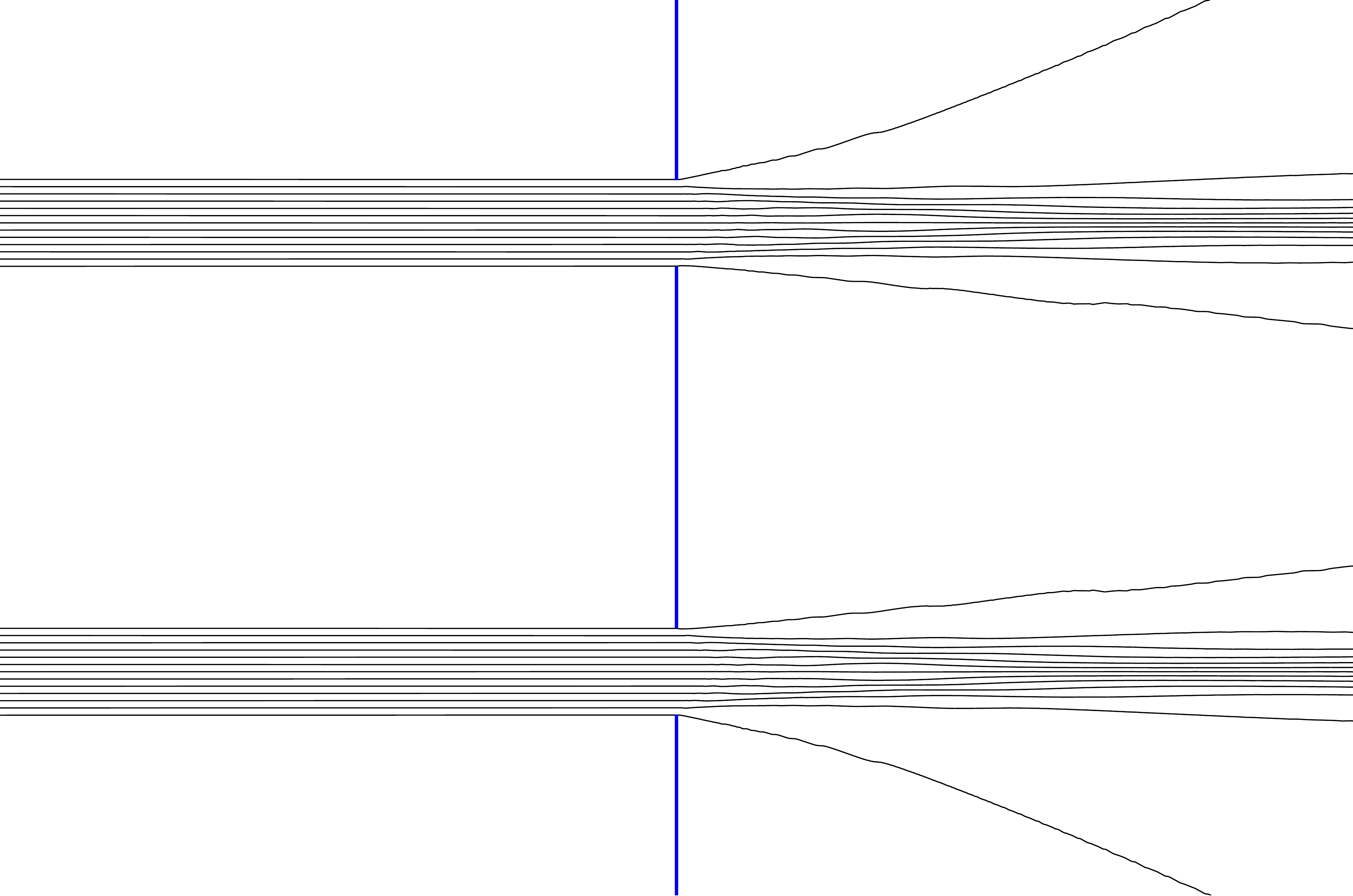}
		\caption[]{{\small $\hbar/100$}}    
		%\label{fig:mean and std of net24}
	\end{subfigure}
	\vskip\baselineskip
	\begin{subfigure}[b]{0.48\textwidth}   
		\centering 
		\includegraphics[width=\textwidth]{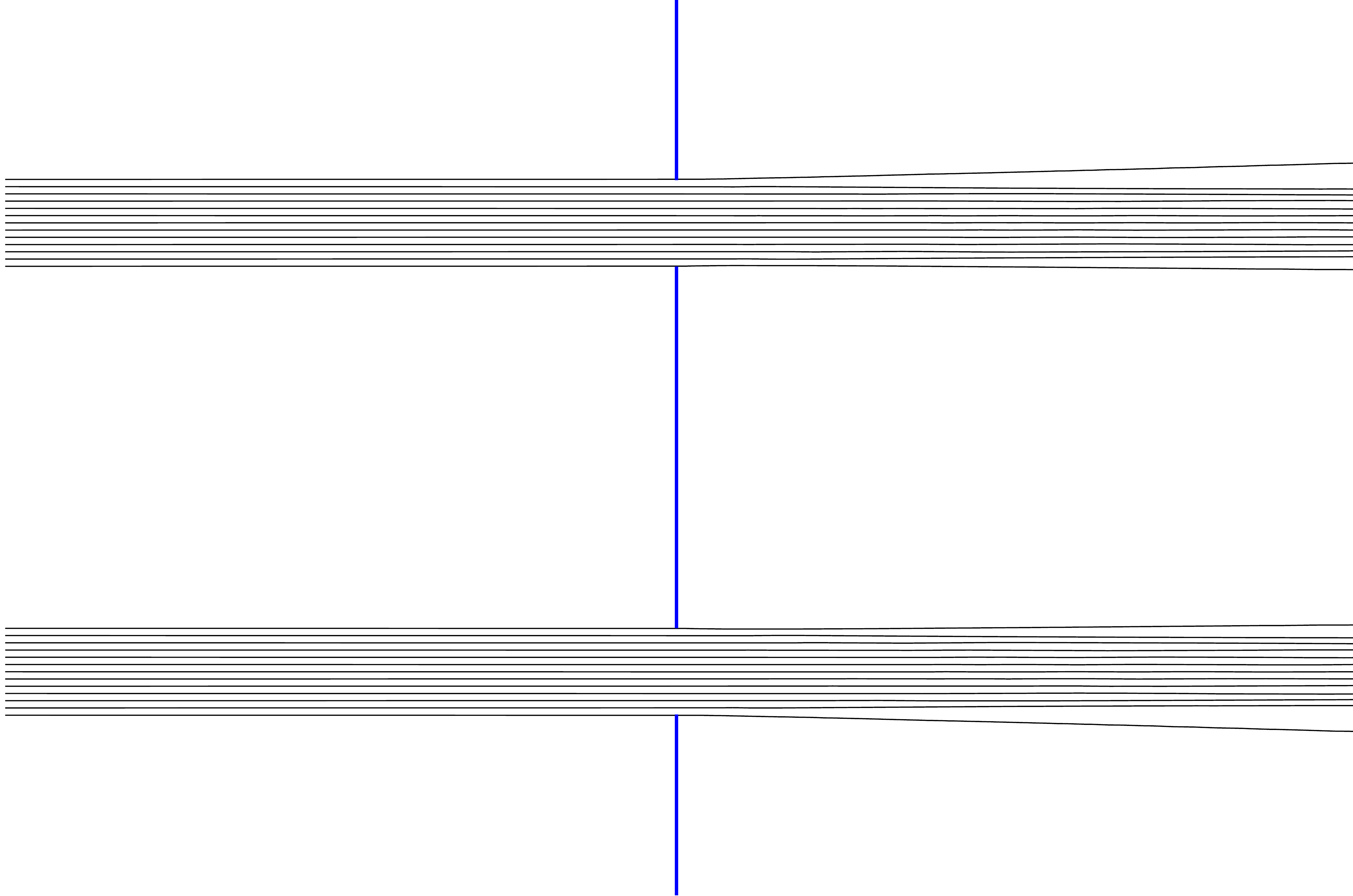}
		\caption[]{{\small $\hbar/1000$}}    
		%\label{fig:mean and std of net34}
	\end{subfigure}
	\quad
	%\hfill
	\begin{subfigure}[b]{0.48\textwidth}   
		\centering 
		\includegraphics[width=\textwidth]{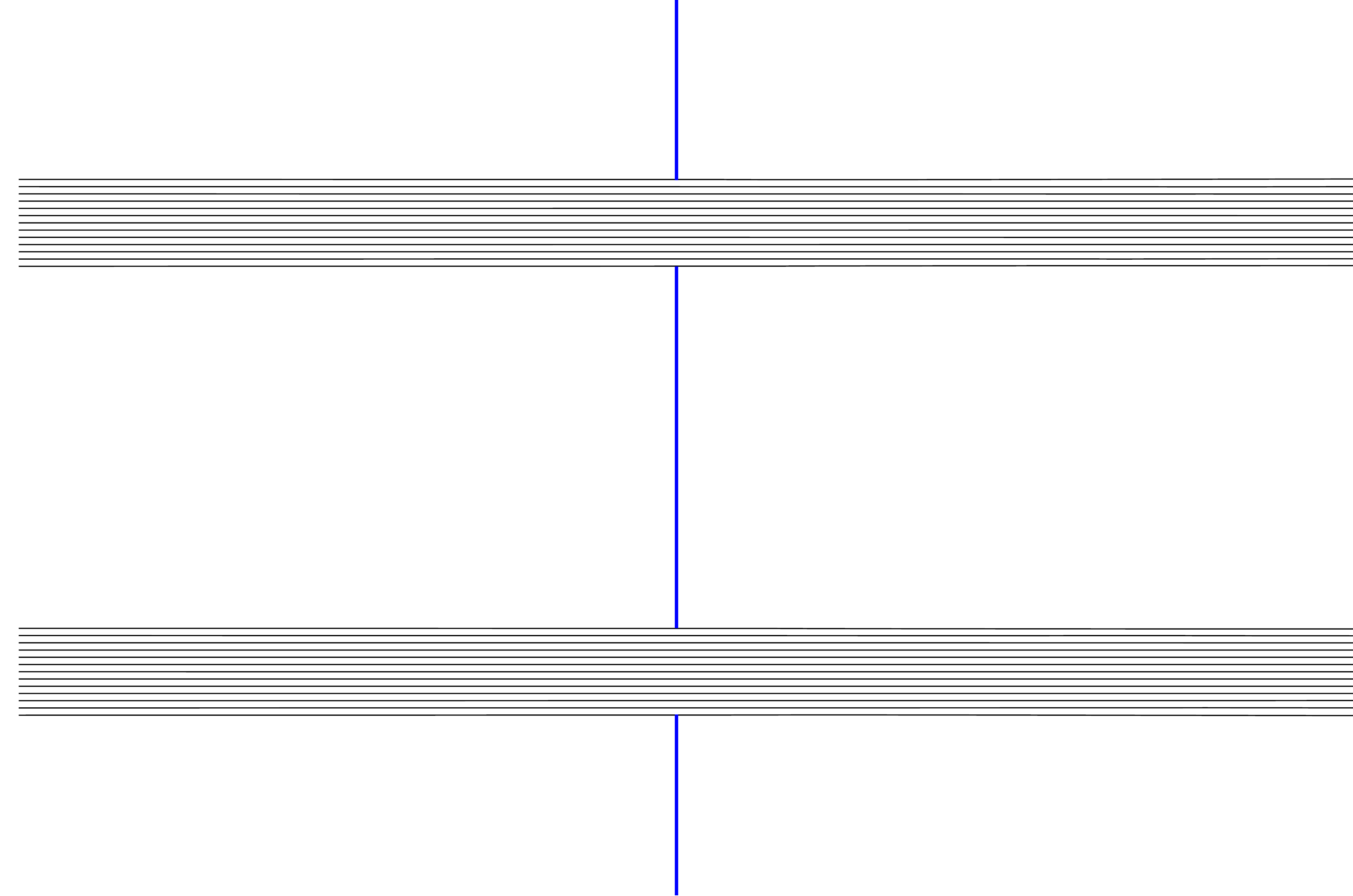}
		\caption[]{{\small $\hbar/10000$}}    
		%\label{fig:mean and std of net44}
	\end{subfigure}\setcounter{figure}{3}
	\caption{\small\label{fig:traj-Young-converg} Convergence of 26 electron trajectories passing through the double-slits when $\hbar$ is divided by 10, 100, 1 000 and 10 000.} 
\end{figure}

Figure~\ref{fig:traj-Young-converg} shows the evolution of 26 of these 100 trajectories when 
the Planck constant is divided by 10, 100, 1000 and 10000 respectively. We see a  continuity between classical mechanics and quantum mechanics:
when h tends to 0, we obtain the convergence of quantum trajectories to classical trajectories.

\begin{figure}[h]
 \centering
\includegraphics[width=0.8\linewidth]{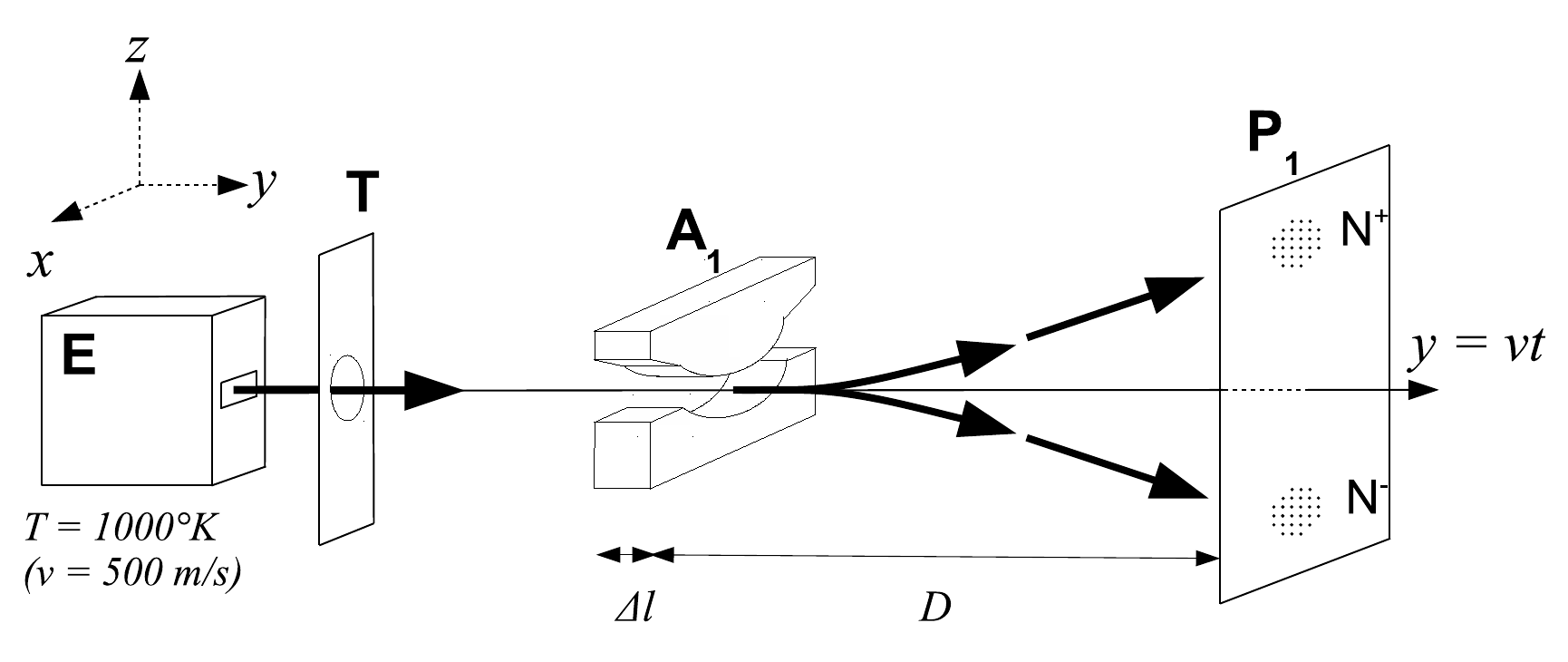}
\caption{\label{fig:figSetG} Schematic configuration of the Stern-Gerlach experiment.}
\end{figure}

In the Stern-Gerlach experiment with a beam of silver atoms (Figure~\ref{fig:figSetG}), the spin of the particle (along the z axis) is not measured  directly; the spin is deduced from the measure of the position of particle impact: spin +1/2 if the impact is in $N^+$, spin -1/2 if the impact is in $N^-$.
%\begin{figure}[h]
%\includegraphics[width=0.6\linewidth]{fig11SG}
%\caption{\label{fig:impacts-SetG}1000 silver atom impacts on the detector $P_1$.}
%\end{figure}
In all the quantum mechanics textbooks~\cite{Feynman1965, Cohen1977, Sakurai2011}, the usual interpretation of this result is based on the three postulates of the measurement (quantization postulate, Born posulate, reduction of the wave function postulate) from the initial wave function (preparation condition) corresponding to a two-component spinor:
\begin{equation}\label{eq:psi0_ses}
    \Psi_{0} = \Psi(t=0) = \left( \begin{array}{c}\cos \frac{\theta_0}{2}e^{ i\frac{\varphi_0}{2}}
                                   \\
                                  \sin\frac{\theta_0}{2}e^{- i\frac{\varphi_0}{2}}
                  \end{array}
           \right)
\end{equation}

In the de Broglie-Bohm weak interpretation, we obtain the same results but without using the three postulates. To do this, we solve the Pauli equation, but with another initial function:
\begin{equation}\label{eq:psi0_aes}
    \Psi_{0}(x,z) = (2\pi\sigma_{0}^{2})^{-\frac{1}{2}}
                      e^{-\frac{x^2+z^2}{4\sigma_0^2}}
                      \left( \begin{array}{c}\cos \frac{\theta_0}{2}e^{ i\frac{\varphi_0}{2}}
                                   \\
                                  \sin\frac{\theta_0}{2}e^{-i\frac{\varphi_0}{2}}
                  \end{array}
           \right)
\end{equation}
which has a spatial extension.

This spatial extension is essential because the two spinor components $\Psi^+(z,t) $ and $\Psi^- (z,t) $ have different evolutions \cite{Gondran2012d} and it is this spatial extension that explains the creation of the two spots $N^+$ and $N^-$. The addition of the particle position then explains the decoherence \cite{Gondran2012d} and proves
the three measurement postulates \cite{Gondran2016}.

The spatial extension of the spinor
(\ref{eq:psi0_aes}) takes into account the particle's
initial position $z_0$ and introduces the Broglie-Bohm
trajectories
\cite{Bohm1952,Holland1993,Dewdney1986}
which is the natural assumption to explain the individual
impacts.

\begin{figure}[h]
\centering
\includegraphics[width=0.8\linewidth]{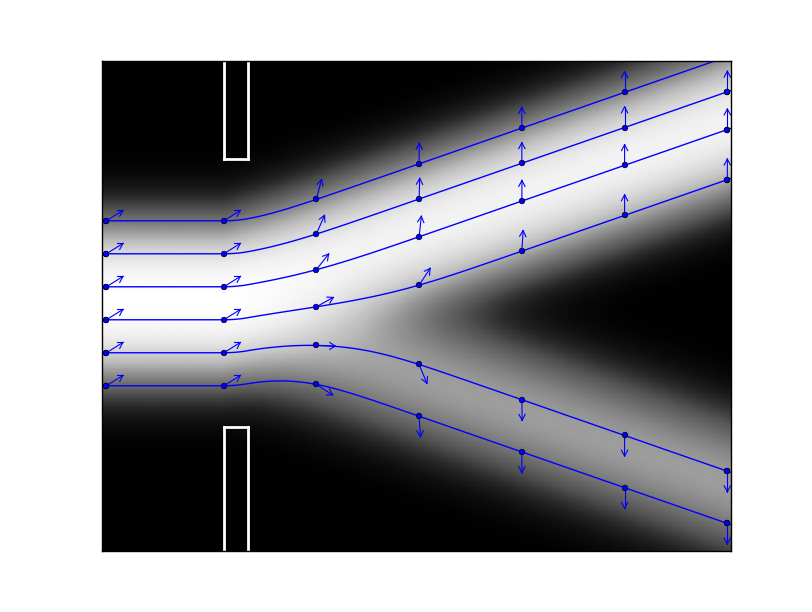}
\caption{\label{fig:traj-SetG} Stern and Gerlach experiment: probability density and ten  silver atom trajectories with
initial spin orientation $(\theta_0=\frac{\pi}{3})$ and initial
position $z_0$; arrows represent the spin vector orientation
$\theta(z,t)$ along the trajectories.
The position of the particle exists before measurement;
the particle then follows a deterministic trajectory and the impact on the screen reveals its position. The atom is both a wave and a particle.}
\end{figure}

Figure~\ref{fig:traj-SetG} presents, for silver atoms with an
initial spinor orientation $(\theta_0=\frac{\pi}{3},\varphi_0=0)$, a plot in the $(Oyz)$ plane of the probability density of the particles and of a set of 10 trajectories whose initial
position $z_0$ has been randomly chosen. The spin
orientations $\theta(z,t)$ are represented by arrows. The numerical values for the simulation are taken from the Cohen-Tannoudji textbook~\cite{Cohen1977}.

Finally, the de
Broglie-Bohm trajectories propose a clear interpretation of the spin 
measurement in quantum mechanics. There is interaction with the
measuring apparatus as is generally stated; and there is indeed a
minimum time required to measure. However this measurement and
this time do not have the signification that is usually applied to them.
The result of the Stern-Gerlach experiment is not the measure of
the spin projection along the $z$-axis, but the orientation of the
spin either in the direction of the magnetic field gradient, or in
the opposite direction. It depends on the position of the particle
in the wave function. We have therefore a simple explanation for
the non-compatibility of spin measurements along different axes.
The measurement duration is then the time necessary for the
particle to point its spin in the final direction. The "measured" value (the spin) is not a pre-existing value such as the mass and the charge of the particle but a contextual value conforming to the Kochen and Specker theorem~\cite{Kochen1967}.

\section{Interpretation of the EPR-B experiment}
\label{sect:EPRB}

We have shown in~\cite{Gondran2016} that the dBB (weak) interpretation could apply to  the EPR-B experiment, the Bohm version of the Einstein-Podolsky-Rosen experiment. To demonstrate this, we have shown that it is possible to replace the singlet spinor of the EPR-B experiment in the configuration space with two single-particle spinors in physical space.

\begin{figure}
 \begin{center}
\includegraphics[width=0.8\textwidth]{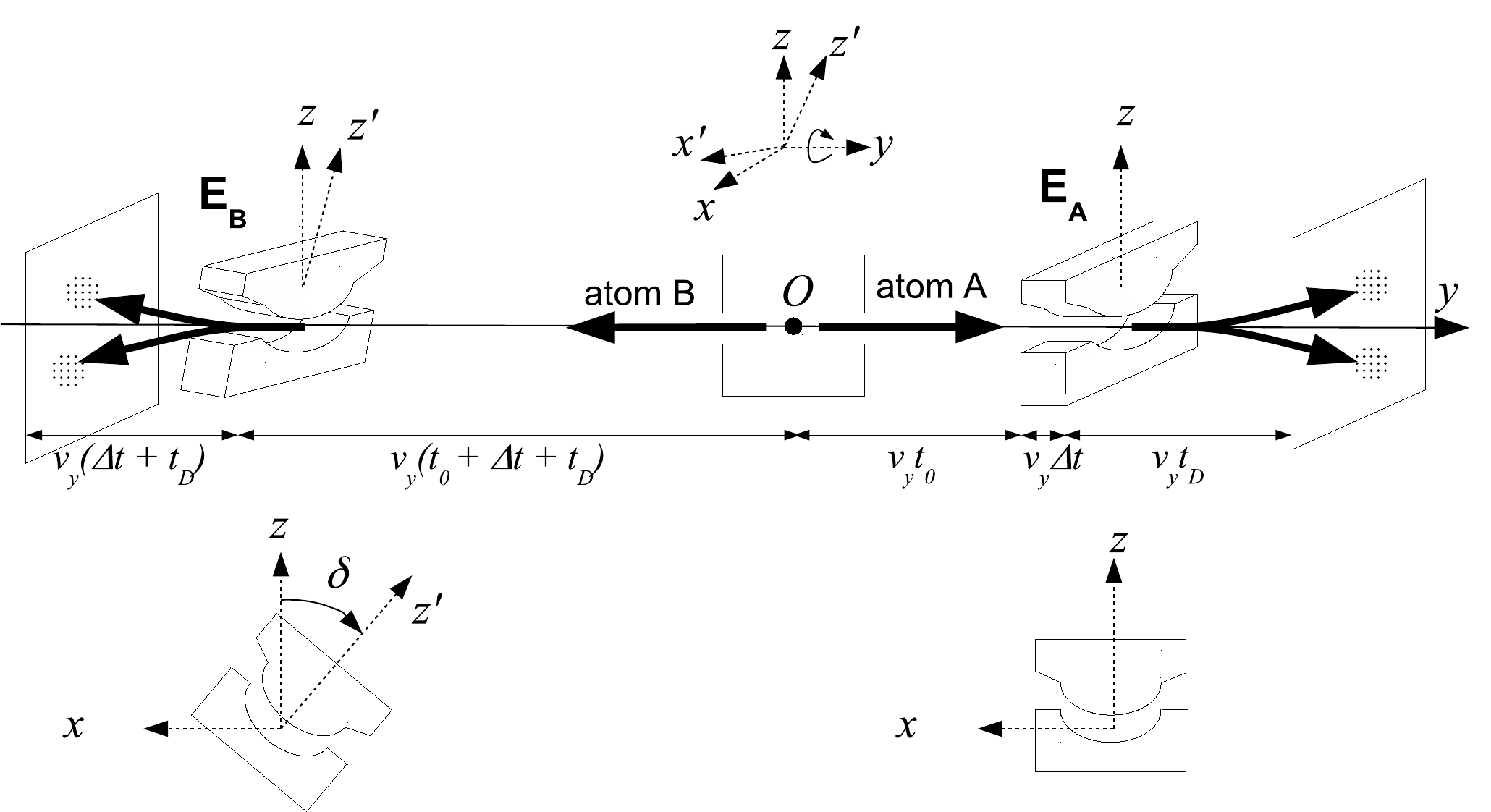}
\caption{\label{fig:EPRBschema} Schematic configuration of the EPR-B experiment: a pair of silver atom entangled by their spin passes through them in turn magnets oriented in offset directions of $\delta$.}
 \end{center}
\end{figure}

Figure~\ref{fig:EPRBschema} shows the EPR-B experiment.
A source $S$ creates in $O$ a pair of identical A and B atoms, entangled by their spin. Atoms A and B separate along the $Oy$ axis in opposite directions 
and encounter two identical Stern-Gerlach type devices $\mathbf{E_A}$ and $\mathbf {E_B}$.

The magnet $\mathbf{E_A}$ measures (or rather straightens) the spin of A in the $Oz$ direction  and the magnet $\mathbf {E_B} $ "measures" the spin of B in the $ Oz'$ direction, direction obtained after a $\delta$ rotation around $Oy$ .
The two "measurements" can be simultaneous or carried out one after the other (here A before B).

To perform this demonstration and to understand the EPR-B experiment in the dBB  interpretation, we solve the Pauli equation in the configuration space for the two entangled particles with initial conditions adapted to the problem.

The initial wave function of the quantum system composed of the two entangled particles is usually the singlet spinor:
\begin{equation}\label{eq:singulet_ses}
    \Psi_{0} =\frac{1}{\sqrt{2}}(|+_{A}\rangle |-_{B}\rangle - |-_{A}\rangle | +_{B}\rangle)
\end{equation}
where $|\pm_{A}\rangle$ (resp. $|\pm_{B}\rangle$) are the eigenvectors of the spin operator $S_{z_A}$ (resp. $S_{z_B}$) in the $Oz$ direction relatif to the A particle  
(resp. B).
In reality, the initial singlet wave function, as in the Stern-Gerlach case, has a spatial extension:% $\Psi_{0}(\textbf{r}_A,\textbf{r}_B)\Psi_{0}$ oÃ¹ $\textbf{r}= (x,z)$ et $f(\textbf{r})=(2\pi\sigma_{0}^{2})^{-\frac{1}{2}} e^{-\frac{x^2 + z^2}{4\sigma_0^2}}$.
\begin{equation}\label{eq:singulet_aes}
    \Psi_{0}(\textbf{r}_A,\textbf{r}_B) =\frac{1}{\sqrt{2}}f(\textbf{r}_A) f(\textbf{r}_B)(|+_{A}\rangle |-_{B}\rangle - |-_{A}\rangle | +_{B}\rangle)
\end{equation}
where $\textbf{r}= (x,z)$ and $f(\textbf{r})=(2\pi\sigma_{0}^{2})^{-\frac{1}{2}}
 e^{-\frac{x^2 + z^2}{4\sigma_0^2}}$.

It is possible to obtain this singlet function~(\ref{eq:singulet_aes}) from the principle of Pauli. To do so, we assume that at the moment of the creation of the two entangled particles A and B, each of the particles has an initial wave function
 $\Psi_0^A(\textbf{r}_A, \theta^A_0,
\varphi^A_0)$ et $\Psi_0^B(\textbf{r}_B, \theta^B_0, \varphi^B_0)$ similar to (\ref{eq:psi0_aes}) but with opposite spins: $\theta_0^B= \pi-\theta_0^A$, $\varphi_0^B= \varphi_0^A -\pi$.

If we then apply the Pauli principle, which states that the wave function of a two-body system must be antisymmetric, we must write: 
\begin{eqnarray}\nonumber
 \Psi_0(\textbf{r}_A,\theta_A, \varphi_A,\textbf{r}_B,\theta_B, \varphi_B)&=&
 \Psi^0_A(\textbf{r}_A,\theta_A, \varphi_A)\Psi^0_B(\textbf{r}_B,\theta_B,
  \varphi_B)\\-\Psi^0_A(\textbf{r}_B,\theta_B, \varphi_B)\Psi^0_B(\textbf{r}_A,\theta_A, \varphi_A)\\
&=&- e^{i \varphi_A} f(\textbf{r}_A)f(\textbf{r}_B)\\(|+_{A}\rangle
|-_{B}\rangle - |-_{A}\rangle|+_{B}\rangle)
\end{eqnarray}

which is the singlet state with spatial extension (\ref{eq:singulet_aes})~\cite{Gondran2016}. 
Again this spatial extension is essential to correctly solve the Pauli equation in space. Moreover, in the dBB theory, the spatial extension is necessary to take into account the position of the atom in its wave function.

    \begin{figure}[h!]
    \centering
   \includegraphics[width=.8\textwidth]{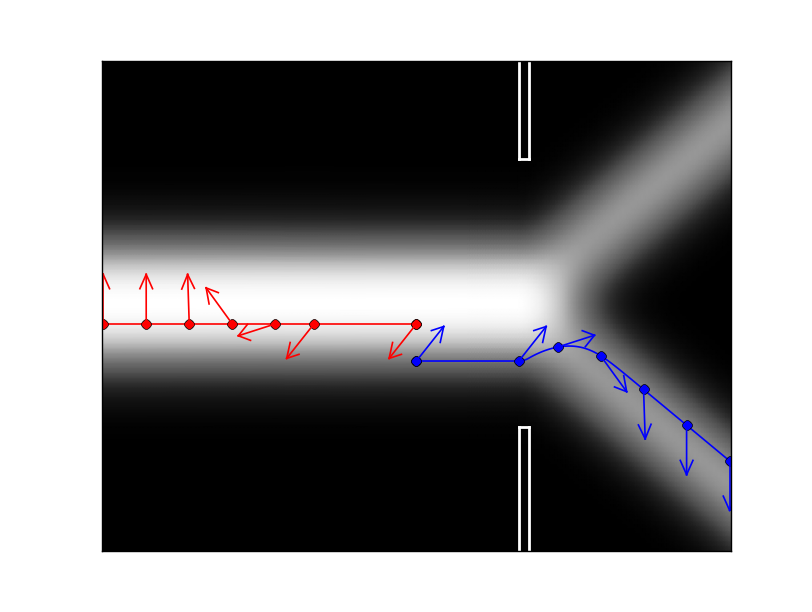}
    \caption{"Measure" of the spin of A: In the center, the two entangled atoms A (blue) and B (red) are created with opposite spins. The particle A goes to the right and crosses the magnet $\mathbf{E_A}$; it behaves in the same way as if it were not entangled. The particle B goes to the left and, during the measurement of A, its spin recovers to remain in opposition to that of A.}
\label{fig:EPR_ddp_traj}
    \end{figure}

In figure~\ref{fig:EPR_ddp_traj}, a pair of entangled atoms is observed during the "measurement" of A by a Stern-Gerlach apparatus $\mathbf{E_A}$.
We show mathematically~\cite{Gondran2016} that the "measured" particle (A) behaves in the device $\mathbf{E_A}$ in the same way as if it were not entangled.
During the "measurement" of A, the density of particle B also evolves as if it were not entangled~\cite{Gondran2016}.    
    
It will be possible to test these two properties experimentally when the EPR-B experiment with atoms is realized.

The particle B goes to the left and, during the "measurement" of A, its spin recovers to always be in opposition to that of the particle A~\cite{Gondran2016}.

When the particle B enters the second magnet $\mathbf{E_B} $ the orientation of its spin $\theta$ with respect to $Oz'$ is $\delta$ if the spin of A has been "measured" down and it is $\pi- \delta$ if the spin of A has been "measured" up.

The second measure (spin of B) is again a Stern-Gerlach "measure". It then perfectly matches the results of quantum mechanics and the violation of Bel's inequalities.

The proposed solution was obtained by solving the two-body Pauli equation that couples spin and spatial degrees of freedom. Moreover, the  measurement postulates of quantum mechanics are not used and can be demonstrated~\cite{Gondran2016}.

In this interpretation, the quantum particle has a local position
like a classical particle, but it also has a non-local behaviour
through the singlet wave function. 

Moreover, the non-local influence in the EPR-B
experiment only concerns the spin orientation, and not the motion
of the particles themselves. This is a key point in the search for
a physical explanation to non-local influence.

The reality of non-locality, that is to say the existence of a supraluminous distance action, has been validated on entangled photons by the Alain Aspect experiments~\cite{Aspect1982b} and his successors Nicolas Gisin~\cite{Tittel1998a} and Anton Zeilinger~\cite{Weihs1998}.
This action contradicts Einstein's [1905] interpretation of restricted relativity.

Should we abandon this interpretation in favor of that of Lorentz-Poincar\'e as suggested by Karl Popper~\cite{Popper1982} p.25~?
\begin{quotation}
 \textit{"Aspect's experiment would then be the first crucial experiment to decide between Lorentz's and Einstein's interpretations of the Lorentz transformations"}
\end{quotation}

The dBB weak theory shows that it is possible to consider quantum mechanics as deterministic and thus show that it is possible to make quantum mechanics and general relativity compatible.
Rehabilitating the existence of a preferential frame of reference, such as that of Lorentz-Poincar\'e and Einstein's ether~\cite{Einstein1920} of 1920, is perhaps the way to reconcile these two theories.

\section{Two possible solutions to complete the dBB weak theory}

We have shown that there are good scientific reasons, both theoretically and experimentally to propose the de Broglie-Bohm interterpretation  for particles in unbound states whose the wave function is in the 3D physical space (particle beam). This is  the de Broglie-Bohm weak interpretation.

In this section, we suggest some ways in which to complete this de Broglie-Bohm weak  interpretation for particles in bound states whose wave function is defined in a 3N-dimensional configuration space.
We retain and discuss two of the solutions proposed in 1927 Solvay Conference:

- the Broglie-Bohm interpretation, which we consider as the de Broglie-Bohm strong interpretation, 

- the soliton wave of the Schr\"odinger interpretation.

\subsection{The de Broglie-Bohm 
strong interpretation}
%\label{sect:}

Continuity and simplicity are the main reasons  to complete the de Broglie-Bohm interpretation restricted to single-particle wave in the 3D physical space in all quantum systems, particularly a quantum system in a configuration space. This is the de Broglie-Bohm strong interpretation. 

In addition, a new result partialy cancels the criticism concerning the configuration space.
For spinless non-relativistic particles, Norsen \cite{Norsen2010}, Norsen, Marian and Oriols \cite{Norsen2014} show that in the de Broglie-Bohm interpretation it is possible to replace the wave function in a 3N-dimensional configuration space by N single-particle wave functions in physical space. These N wave functions in 3D-space are the N \textit{conditional wave functions} of a subsystem introduced by D\"urr, Goldstein and Zanghi \cite{Durr1992,Durr2004}. We have shown \cite{Gondran2016} and in the previous section recall that this replacment of the wave function in the configuration space by single-particle functions in the 3D-space is also possible for particles with spin, in particular for the particles of the EPR-B experiment.

\subsection{Schr\"odinger's soliton wave}
%\label{sect:}

As early as March 1926, Schr\"odinger proposed that the wave be considered as the only reality, particles being a consequence. Indeed, for a harmonic oscillator, he showed~\cite{Schrodinger1926} (Schr\"odinger,1926) there is a wave packet which " \textit{remains compact, and does not spread out into larger regions as time goes on, as we were accustomed to make it to, for example in optics}".
They are the coherent states that currently have a great role in quantum optics and in
second quantization~\cite{Glauber1965}. This type of solution, which remains spatially well localized and does not disperse with time is also called a soliton. Schr\"odinger (1926) anticipated that a soliton solution must exist for the electron of the hydrogen atom : " \textit{We can definitely foresee that, in a similar way, wave groups can be constructed which move around highly quantized Kepler ellipses and the representation by wave mechanics of the hydrogen electron. But the technical dificulties in the calculation are greater than in particularly simple example we have treated here.}" This soliton solution for the electron of the hydrogen atom remained a dream until recently. A mathematical solution was found in 1994 ~\cite{Bialynicki1994, Buchleitner1995, Buchleitner2002}(Bialynicki et al., 1994 ; Delande, 1995),  and
discovered recently in 2004 by Maeda and Gallagher~\cite{Maeda2004} on
Rydberg atoms.

\begin{figure}[h]
\centering
\includegraphics[width=0.5\linewidth]{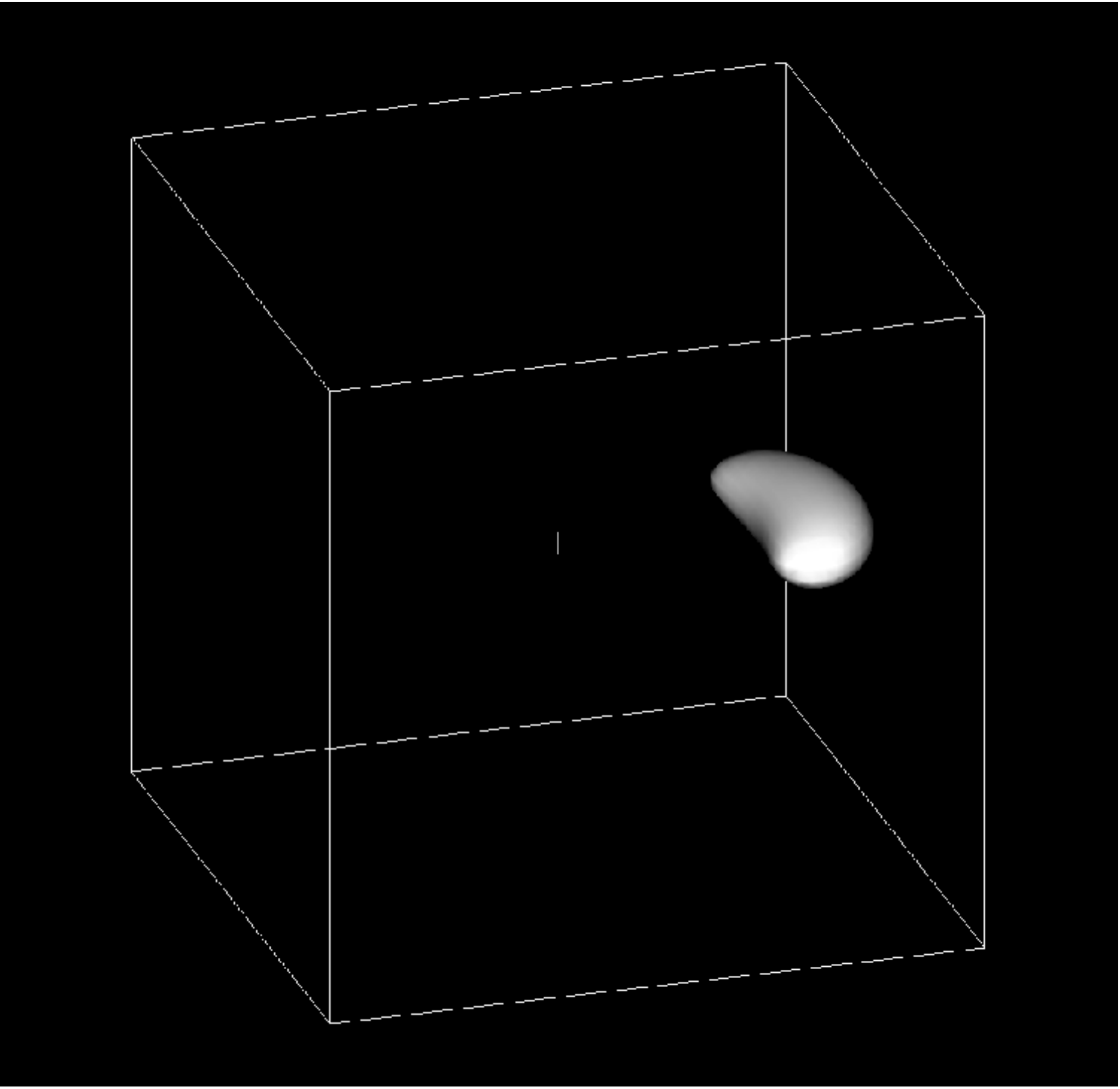}
\caption{\label{fig:traj-Delande}  Coherent wave packets of the hydrogen atom (Buchleitner et Delande, 1995)~\cite{Buchleitner1995}.}
\end{figure}

Figure~\ref{fig:traj-Delande} shows a wave packet in the form of a banana calculated in a frequency field that is close to 30 GHz with a principal quantum number and equal to 60. The packet is approximately four thousand Bohr radii from the nucleus and revolves around it in the horizontal plane without changing shape.~\cite{Buchleitner1995}

Let us consider the following preparation of the particles.

\begin{Definition}[Recognizable quantum particle]\label{defRecognizableParticle}
- A quantum particle is said to be recognizable
prepared if its initial probability density $\rho^{\hbar}_{0}(\mathbf{x})$
and its initial action $S^{\hbar}_{0}(\mathbf{x})$ converge, when $\hbar\to 0$, respectively to a Dirac distribution $\delta(\textbf{x}-\textbf{x}_0)$ and to a regular function
$S_{0}(\mathbf{x})$. 
\end{Definition}

It is the case of the coherent state of the two dimensional harmonic oscillator in the field $V(\textbf{x})=\frac{1}{2}m \omega^{2}\textbf{x}^{2}$.

The coherent states are built~\cite{Cohen1977} from the initial wave
function $\Psi_{0}(\textbf{x})$, which corresponds to the density
and initial action $ \rho^{\hbar}_{0}(\mathbf{x})= ( 2\pi \sigma
_{\hbar}^{2}) ^{-1} e^{-\frac{( \textbf{x}-\textbf{x}_{0})
^{2}}{2\sigma _{\hbar}^{2}}}$ and $S^{\hbar}_{0}(\mathbf{x})= m \textbf{v}_{0}\cdot
\textbf{x}$ with $ \sigma_\hbar=\sqrt{\frac{\hbar}{2 m \omega}}$. We
verify that, when $\hbar\to 0$, $ \rho^{\hbar}_{0}(\mathbf{x})$ converges to the Dirac distribution $\delta(\textbf{x}-\textbf{x}_0)$ and that $ S_{0}(\mathbf{x})=S^{\hbar}_{0}(\mathbf{x})$ is a regular function. With these initial conditions, the density
$\rho^{\hbar}(\textbf{x},t)$ and the action
$S^{\hbar}(\textbf{x},t)$, solutions to the Madelung equations
(\ref{eq:Madelung1})(\ref{eq:Madelung2})(\ref{eq:Madelung3}), are
equal to ~\cite{Cohen1977}:
$\rho^{\hbar}(\textbf{x},t)=\left( 2\pi \sigma_{\hbar} ^{2}
\right) ^{-1}e^{- \frac{( \textbf{x}-\xi(t)) ^{2}}{2\sigma_{\hbar}
^{2} }}$ and $S^{\hbar}(\textbf{x},t)= + m \frac{d\xi
(t)}{dt}\cdot \textbf{x} + g(t) - \hbar\omega t$, where $\xi(t)$
is the trajectory of a classical particle evolving in the
potential $V(\textbf{x})=\frac{1}{2} m \omega^{2} \textbf{x}^2 $,
with $\textbf{x}_0$ and $\textbf{v}_0$ as initial position and
velocity and $g(t)=\int _0 ^t ( -\frac{1}{2} m (\frac{d\xi
(s)}{ds})^{2} + \frac{1}{2} m \omega^{2} \xi(s)^2) ds$.

\begin{Theoreme}\label{t-convergenceparticulediscerne}~\cite{Gondran2011a,Gondran2012a}- 
For the harmonic oscillator, when $\hbar\to 0$,
the density $\rho^{\hbar}(\textbf{x},t)$ and the action
$S^{\hbar}(\textbf{x},t)$ converge to
\begin{equation}
\rho(\textbf{x},t)=\delta( \textbf{x}- \xi(t)) ~\text{ and }~
S(\textbf{x},t)= m \frac{d\xi (t)}{dt}\cdot\textbf{x} + g(t)
\end{equation}
where $S(\textbf{x},t)$ and the trajectory $\xi(t)$ are solutions
to the deterministic Hamilton-Jacobi equations: 
\begin{eqnarray}
\frac{\partial S\left(\textbf{x},t\right) }{\partial
t}|_{\textbf{x}=\xi(t)}
+\frac{1}{2m}(\nabla S(\textbf{x},t))^{2}|_{\textbf{x}=\xi(t)}
+V(\textbf{x})|_{\textbf{x}=\xi(t)}=0\label{eq:statHJponctuelle1b}
\end{eqnarray}
\begin{eqnarray}\label{eq:statHJponctuelle1c}
\frac{d\xi(t)}{dt}=\frac{\nabla S(\xi(t),t)}{m}\\
\label{eq:statHJponctuelle1d}
S(\textbf{x},0)= m \textbf{v}_0 \textbf{x}
~\text{ and }~\xi(0)=\textbf{x}_0.
\end{eqnarray}
\end{Theoreme}

Therefore, the kinematic of the wave packet converges to the
single harmonic oscillator described by $\xi(t)$, which corresponds to \textit{a
classical particle} for which we know the position and the velocity. It is then possible to consider, unlike for the semi-classical statistical particles, that the wave function can be
viewed as a single quantum particle. Then, we consider this deterministic quantum particle as the classical particle. 
The \textit{semi-classical deterministic quantum particle} is in line with the Copenhagen interpretation
of the wave function, which contains all the information on the
particle. A natural interpretation was proposed by
Schr\"odinger~\cite{Schrodinger1926} in 1926 for the coherent
states of the harmonic oscillator: the quantum particle is a
spatially extended particle, represented by a wave packet whose
center follows a classical trajectory. In this interpretation, 
the first two usual postulates of quantum mechanics are maintained. 
The others are not necessary. 
Then, the particle center is the mean value of the position 
($X(t)=\int x\vert\Psi(\textbf{x},t) \vert^2 dx$)  and satisfies the Ehrenfest theorem~\cite{Ehrenfest1927}.
 
%We postulate that the quantum velocity field is now $ \mathbf{v}^{\hbar}\left( \mathbf{x},t\right) =\frac{\mathbf{\nabla }S^{\hbar}\left( \mathbf{%
%x,}t\right) }{m} +\frac{\hbar}{2 m}\triangledown ln\rho^{\hbar} (\textbf{x},t) \star \textbf{k}$ 
%where $\textbf{k}$ is the unit vector parallel to the spin. For the coherent state of the 2D harmonic oscillator, this velocity field is equal to $\mathbf{v}^{\hbar}\left( \mathbf{x},t\right) = \xi(t)+ \omega\textbf{k}\star (\textbf{x}-\xi(t))$. Then, the extended particle (soliton) has the same motion as a 2D spinning top.  

This approach is compatible with the construction of quantum field theory in which the first stage is to consider free fields with point-like particles, whereas the second stage is to consider interacting fields in which the (dressed) particle can no longer be considered as point-like. 
It is this approach that joins de Broglie's theory of the double solution that we will develop in a future article.

\section{Conclusion}
\label{sect:conclusion}
It was well known that the energy spectrum of quantum particles in unbounded states was continuous, like the energy spectrum of classical particles in the same field.

We have shown that we can go much further and that there exists a mathematical continuity between these classical and quantum particles prepared in the same way. There is indeed a convergence between the equations of quantum mechanics (the Madelung equations representing the Schr\"odinger equation) and the equations of classical mechanics (the statistical Hamilton-Jacobi equations) when the Planck constant is set to 0. But to bring about this convergence, we must specify the initial conditions of the density and the phase of the wave function, ie the preparation of the quantum particles. This was only possible for a restricted class of quantum particles, the unrecognizable quantum particles, which converge towards a class of restricted classical particles, classical unrecognizable particles.

This continuity between quantum mechanics and classical mechanics is reinforced by the Minplus path integral which is the analog in classical mechanics of Feynman's path integral in quantum mechanics. This distinction between the Hamilton-Jacobi and Euler-Lagrange actions, based on the Minplus path integral, makes it easier to understand the principle of least action.

Furthermore, the introduction into classical mechanics of the concepts of unrecognizable 
particles verifying the statistical Hamilton-Jacobi equations can provide a 
simple answer to the Gibbs paradox of classical statistical mechanics. 

Finally, since the Hamilton-Jacobi action drives the classical unrecognizable particles, we have assumed that the wave function also drives unrecognizable quantum particles. This is the dBB  weak interpretation.

For the other quantum systems prepared differently, in particular the quantum particles in bound states whose wave function is defined in a 3N-dimensional configuration space, we propose two possible solutions on how to complete the de Broglie-Bohm weak interpretation.

This interpretation of quantum mechanics following the preparation of the system sheds light on the discussions between the founding fathers, in particular the discussion of the Solvay Congress of 1927. 
Indeed, one may consider that the misunderstanding between them may have come from the fact that they each had an element of truth: Louis de Broglie's pilot-wave interpretation for the unrecognizable particles, Schr\"odinger's soliton interpretation for the harmonic oscillator and Born's statistical interpretation for 
the diffusion states. But each applied his particular case to the general case and they consequently 
made mutually incompatible interpretations.

\bibliographystyle{elsarticle-num}
\bibliography{biblio_mq}

\end{document}